\documentclass[reprint,onecolumn,showpacs,nofootinbib,prd,amssymb]{revtex4}
\usepackage{latexsym}
\usepackage{graphicx,epsf, epsfig, amssymb, amsmath}

\usepackage[usenames,dvipsnames]{xcolor}
\newcommand{\aap}{Astron.\ Astrophys.}
\newcommand{\mnras}{Mon.\ Not.\ R.\ Astron.\ Soc.}

\newcommand{\apjl}{Astrophys.\ J.\ Lett.}
\newcommand{\apjs}{Astrophys.\ J.\ Suppl.\ Ser.}

\begin{document}

\title{Relativistic formulation 
of the Hall-Vinen-Bekarevich-Khalatnikov superfluid hydrodynamics}

\author{
M.~E.~Gusakov$^{1,2}$} \affiliation{$^1$Ioffe Physical-Technical Institute of the Russian Academy
  of Sciences, Polytekhnicheskaya 26, 194021 Saint-Petersburg, Russia}
\affiliation{$^2$Saint-Petersburg State Polytechnical University, Polytekhnicheskaya
  29, 195251 Saint-Petersburg, Russia}

\begin{abstract} 
The relativistic analogue of the Hall-Vinen-Bekarevich-Khalatnikov (HVBK) hydrodynamics
is derived making use of the phenomenological method similar to that used 
by Bekarevich and Khalatnikov \cite{bk61} in their derivation of HVBK-hydrodynamics.
The resulting equations describe 
a finite-temperature superfluid liquid with the distributed vorticity.
The main dissipative effects, including mutual friction, are taken into account.
The proposed hydrodynamics is needed for reliable 
modeling of the
dynamical properties of superfluid neutron stars.
\end{abstract}
\received{16 October 2015}
\accepted{18 February 2016}


%
%
%
%
%
%



\maketitle

\section{Introduction}

Despite the fact that superfluid flow must be irrotational,
it is well known \cite{ll87, khalatnikov00, putterman74, donnelly05} 
that in a rotating bucket a superfluid mimics solid body rotation 
{\it on average} by creating arrays of topological defects -- vortex lines,
near which the irrotationality condition breaks down. 

Hall and Vinen \cite{hv56} developed in 1956 a 
coarse-grained hydrodynamic equations capable of describing 
a superfluid liquid with the continuously distributed vorticity. 
Their equations are only valid in situations when a typical 
length scale of the problem is much larger than the intervortex spacing.
Later in 1960-1961 Hall \cite{hall60} and, independently, 
Bekarevich and Khalatnikov \cite{bk61}
presented a more elaborated version of these equations
which is now called Hall-Vinen-Bekarevich-Khalatnikov (HVBK) hydrodynamics.
Note that the most general phenomenological derivation 
of HVBK-hydrodynamics, based upon conservation laws, were given by 
the last two authors in the abbreviation (Bekarevich and Khalatnikov).
Subsequently, many authors have repeated and analyzed 
their derivation in order to generalize it and/or make it more transparent 
(see Donnelly \cite{donnelly05} and Sonin \cite{sonin87}
for details and, especially, Refs.\ \cite{clark63, holm01}).
The main conclusion of their work is that, basically,
the structure of the HVBK-hydrodynamics remains unaffected if one is not interested in 
the oscillation modes related to the elasticity of the vortex lattice \cite{bc83,cb86}.

The HVBK-hydrodynamics has received a great deal of attention in 
relation to the interpretation of liquid helium II experiments \cite{donnelly05, sonin87}
and, somewhat unexpectedly, in relation to the neutron star physics 
(see, e.g., Refs.\ \cite{lm00,ly03,pmgo06, hap09, agh09}).
Since HVBK-equations are essentially non-relativistic, 
the majority of studies of superfluid neutron-star dynamics 
have been performed in the non-relativistic framework. 
This framework is (as a rule) 
acceptable for a qualitative analysis of the problem but is inadequate
for obtaining the quantitative results since neutron stars 
are essentially relativistic objects.

Clearly, one needs a Lorentz-covariant formulation of HVBK-hydrodynamics.
In the literature there were only few attempts 
to find such a 
formulation \cite{lk82, cl95} 
(see also Ref.\ \cite{kl82}, lectures \cite{carter00}, and references therein).
The authors of these works restrict themselves to the case of
a vanishing temperature ($T=0$), when there are no thermal excitations (normal component)
in the liquid and hence no dissipative interaction (the so called ``mutual friction'')
between the superfluid and normal liquid components.
The resulting hydrodynamics,
generalized subsequently to describe superfluid mixtures \cite{cl98, lsc98},
 have then been applied 
to model oscillations of cold ($T=0$) superfluid rotating neutron stars
in Ref.\ \cite{yl03}. 
Note, however, that in many physically interesting situations 
the approximation of vanishing stellar temperature
is not justified and leads to {\it qualitatively} wrong results when studying
neutron star dynamics
(see, e.g., Refs.\ \cite{kg11,gkcg13,gck14a,gck14b,kg14a,kg14b} 
for illustration of principal importance
of finite temperature effects in some problems).
Moreover, as we argue in Appendix~\ref{incons},
the hydrodynamics 
of Refs.\ \cite{lk82,cl95} 
is internally inconsistent,
which can have important consequences
for those problems 
(see Ref.\ \cite{bgt13} for an example) 
for which 
the contribution of the vortex energy 
to the total energy density cannot be neglected.%
%
\footnote{The results of Ref.\ \cite{yl03} remain unaffected 
since it (legitimately) ignores 
a small vortex contribution to the total energy density.}
%

The aim of the present study is to fill the existing gap 
by deriving the self-consistent relativistic dissipative HVBK-hydrodynamics, 
valid at arbitrary temperature.
Our derivation will closely follow the ideas of the original derivation 
of Bekarevich and Khalatnikov \cite{bk61}.

The paper is organized as follows.
In Sec.\  \ref{Sec2} we present the derivation 
of the well known vortex-free superfluid relativistic hydrodynamics.
In Sec.\ \ref{neglected} we formulate the relativistic HVBK-hydrodynamics
under the assumption that the contribution of vortices to the total energy density
of a superfluid can be neglected.
In Sec.\ \ref{included} this assumption is relaxed 
and the most general relativistic HVBK-equations
are found.
Finally, we conclude in Sec.\ \ref{concl}.

The paper also contains a number of important appendices.
In Appendix \ref{HVBK} we present the original (non-relativistic) HVBK-hydrodynamics;
in Appendix \ref{full_hydro} we list the full system of equations of relativistic HVBK-hydrodynamics;
in Appendix \ref{nonrel_limit} we analyze the non-relativistic limit 
of one of the most important equations
of the proposed hydrodynamics 
-- the superfluid equation;
in Appendix \ref{evortex} we find the vortex contribution to the energy density;
in Appendix \ref{micro} we present an alternative microscopic derivation
of the vortex contribution to the energy-momentum tensor 
(more precisely, derivation of its spatial components);
finally, in Appendix \ref{incons} we discuss the internal inconsistency 
of the zero-temperature vortex hydrodynamics of Refs.\ \cite{lk82,cl95}.
 
Unless otherwise stated, in what follows 
the speed of light $c$, the Planck constant $\hbar$,
and the Boltzmann constant $k_{\rm B}$ are all set to unity,
$c=\hbar=k_{\rm B}=1$.

\section{Relativistic superfluid hydrodynamics in the absence of vortices}
\label{Sec2}

\subsection{General equations}
\label{Sec2a}

Neglecting vortices, 
relativistic superfluid hydrodynamics for a one-component liquid 
has been studied in many papers and is well known 
(see, e.g., \cite{kl82, lk82, ck92b, son01, pd03, gusakov07, peletminskii09, bbm11, hlsy11, amss13, bbmy14, stetina15}). 
Here we present its derivation 
partly in order to establish notations and partly because, as we believe, 
it can be of independent interest.
Our derivation 
adopts the same strategy as that used by Khalatnikov \cite{khalatnikov00} 
to derive equations 
of {\it non-relativistic} superfluid hydrodynamics.

Hydrodynamic equations include the energy-momentum conservation 
\begin{equation}
\partial_{\mu} T^{\mu\nu}=0
\label{Tmunu}
\end{equation}
and particle conservation 
\begin{equation}
\partial_\mu j^{\mu}=0,
\label{jmu}
\end{equation}
where $\partial_\mu \equiv \partial/\partial x^{\mu}$;
$T^{\mu \nu}$ is the energy-momentum tensor 
(which must be symmetric)
and $j^{\mu}$ is the particle four-current density.
Here and below, unless otherwise stated, 
$\mu$, $\nu$, and other Greek letters
are space-time indices 
running over $0$, $1$, $2$, and $3$. 
Generally, $T^{\mu\nu}$ and $j^{\mu}$
can be presented as
\begin{eqnarray}
T^{\mu\nu} &=& \underline{(P+\varepsilon) u^{\mu} u^\nu} + \underline{P g^{\mu\nu}} + \Delta T^{\mu\nu},
\label{Tmunu2}\\
j^{\mu} &=& \underline{n u^{\mu}} + \Delta j^{\mu},
\label{jmu2}
\end{eqnarray}
where $P$ is the pressure given by Eq.\ (\ref{pres}) below; 
$\varepsilon$ is the energy density; 
$n$ is the number density; 
$g_{\mu\nu}={\rm diag}(-1,1,1,1)$ is the space-time metric.%
\footnote{Throughout the paper we assume that the metric is flat.
Generalization of our results to arbitrary $g_{\mu\nu}$ is straightforward
provided that all relevant length scales of the problem 
(e.g., thermal excitation mean free path)
are small enough compared with the 
characteristic gravitational length scale (e.g., neutron star radius) \cite{weinberg72}.
In the latter case general relativity effects can easily be incorporated into hydrodynamics 
by replacing ordinary derivatives in all equations with their covariant analogues.}
Finally, $u^{\mu}$ is the four-velocity of the normal (non-superfluid) liquid component 
(thermal excitations), normalized by the condition
\begin{equation}
u_{\mu} u^{\mu} = -1.
\label{norm1}
\end{equation}
The underlined terms in Eqs.\ (\ref{Tmunu2}) and (\ref{jmu2})
have the familiar form of, respectively, 
the energy-momentum tensor and particle current density
of nonsuperfluid matter (see, e.g., Ref.\ \cite{ll87}). 
Correspondingly, additional ``superfluid'' terms 
$\Delta T^{\mu\nu}$ and $\Delta j^{\mu}$
characterize deviation of superfluid hydrodynamics from the ordinary one.
Note that the thermodynamic quantities introduced in Eqs.\ (\ref{Tmunu2}) and (\ref{jmu2})
do not have any direct physical meaning 
unless a comoving frame 
where they are measured (defined) is specified.
In what follows we {\it define} the comoving frame by the condition 
$u^{\mu}=(1,0,0,0)$ indicating,  
that 
it is the frame where 
the normal liquid component is at rest.
This definition coincides with the definition of the comoving frame
in the ordinary relativistic hydrodynamics.
It means, in particular, 
that the components $T^{00}$ and $j^{0}$ in this frame
are given by the conditions, $T^{00}=\varepsilon$ and $j^{0}=n$,
which, in an arbitrary frame, translates into
\begin{eqnarray}
u_{\mu}u_{\nu} T^{\mu\nu} &=& \varepsilon,
\label{condTmunu} \\
u_{\mu} j^{\mu} &=& -n,
\label{condjmu}
\end{eqnarray}
or, 
in view of the expressions (\ref{Tmunu2}) and (\ref{jmu2}) to
\begin{eqnarray}
u_{\mu}u_{\nu} \, \Delta T^{\mu\nu} &=& 0,
\label{condTmunu2}\\
u_{\mu} \Delta j^{\mu} &=& 0.
\label{condjmu2}
\end{eqnarray}

As a matter of fact, we can extract even more information 
about the form of $j^{\mu}$ in the comoving frame.
Since in that frame normal component does not move, 
spatial components of the current $j^i$ ($i=1$, $2$, $3$) 
are non-zero only because of the motion of superfluid component.
In the non-relativistic limit the contribution of the superfluid
component in this situation would be $\rho_{\rm s} {\pmb V}_{\rm s}$, 
where $\rho_{\rm s}$ is the superfluid density 
and ${\pmb V}_{\rm s}={\pmb \nabla} \phi/m$ is the superfluid velocity.
(Here $m$ is the bare particle mass and $\phi$ is a scalar proportional 
to the phase $\Phi$ of the condensate wave function;
for Bose-Einstein condensate $\phi=\Phi$, 
for Cooper-pair condensate $\phi=\Phi/2$ \cite{ll80,ga06}.)
By analogy, in the relativistic case it is natural 
to introduce a superfluid four-velocity 
\begin{equation}
V_{({\rm s})}^{\mu} \equiv \frac{\partial^{\mu}\phi}{m},
\label{Vs}
\end{equation}
and
assume that 
$j^i$ can be represented as
$j^i =m Y\, V_{({\rm s})}^{i} = Y  \, \partial^i \phi$,
where $Y$ is some coefficient, 
a relativistic equivalent of the
superfluid density $\rho_{\rm s}$
(it is easily verified that 
in the non-relativistic limit
$Y=\rho_{\rm s}/(m^2 c^2)$ in dimensional units \cite{ga06}).
Consequently, in the comoving frame
one has
(see Eqs.\ \ref{jmu2} and \ref{condjmu2})
\begin{eqnarray}
\Delta j^0 &=& 0,
\label{j0}\\
 \Delta j^i &=& j^i = Y  \partial^i \phi.
 \label{ji}
\end{eqnarray}
In an arbitrary frame this expression can be rewritten by introducing a new 
four-vector, $b^{\mu}$, as
\begin{equation}
\Delta j^\mu =Y\,  (\partial^{\mu} \phi +b^{\mu}),
\label{djmu3}
\end{equation}
To satisfy Eq.\ (\ref{ji}), 
a spatial part of $b^{\mu}$ should vanish 
in the comoving frame, 
$b^{i}=0$.
That is, $b^{\mu}$ and $u^{\mu}$ should be collinear in that frame, 
hence they must be collinear in all other frames, i.e.,
\begin{equation}
b^{\mu}=-B u^{\mu},
\label{bmu} 
\end{equation}
where $B$ is some scalar to be determined below.
In view of Eqs.\ (\ref{condjmu2}) and (\ref{djmu3})
$B$ and $\phi$ are interrelated by the following 
equation
\begin{equation}
u_{\mu} \partial^{\mu} \phi =- B.
\label{sfleq}
\end{equation}
Note that, Eq.\ (\ref{j0}) is then automatically satisfied.

Let us now introduce
a new four-vector,
\begin{equation}
w^{\mu} \equiv \partial^{\mu} \phi - B u^{\mu}
\label{new}
\end{equation}
instead of $\partial^{\mu} \phi$.
Since this vector depends on the four-gradient 
of the scalar $\phi$, 
it is not arbitrary and is constrained by the condition
\begin{equation}
\partial_{\mu}(w_{\nu}+ B u_{\nu})=\partial_{\nu}(w_{\mu}+B u_{\mu}),
\label{weq}
\end{equation}
which is simply the statement that 
$\partial_{\mu}\partial_{\nu} \phi 
= \partial_{\nu}\partial_{\mu} \phi$.
In what follows Eq.\ (\ref{weq}) is called the {\it potentiality} condition
or simply the {\it superfluid} equation.
In terms of the new four-vector $w^{\mu}$
one has (see Eqs.\ \ref{jmu2} and \ref{djmu3})
\begin{equation}
j^{\mu}=n u^{\mu} + Y w^{\mu},
\label{jmu4}
\end{equation}
while the condition (\ref{condjmu2}) transforms into
\begin{equation}
u_{\mu} w^{\mu}=0.
\label{wu}
\end{equation}

Eqs.\ (\ref{Tmunu})--(\ref{Tmunu2}), (\ref{condTmunu}), (\ref{condjmu}),
and (\ref{weq})--(\ref{wu}) are key equations that will be used below.
They should be supplemented by the second law of thermodynamics.

In a normal matter the energy density $\varepsilon$ 
of a one-component liquid can generally be presented as a function
of the number density $n$ and the entropy density $S$.
In superfluid matter, there is an additional degree of freedom
associated with the vector $w^{\mu}$.
One can construct two scalars associated with $w^{\mu}$, namely, 
$u_{\mu }w^{\mu}$ and $w_{\mu} w^{\mu}$.
The first scalar vanishes on account of (\ref{wu}), 
so that
$\varepsilon=\varepsilon(n,\, S, \, w_{\mu} w^{\mu})$.
Consequently, variation of $\varepsilon$ can generally be written as
\begin{equation}
d \varepsilon = \mu \, dn + T \, dS + \frac{\Lambda}{2} \,  d \left( w_{\mu} w^{\mu} \right),
\label{2ndlaw}
\end{equation}
where we defined the relativistic chemical potential 
$\mu \equiv \partial \varepsilon(n,\, S,\, w_{\mu} w^{\mu})/\partial n$;
temperature $T \equiv \partial \varepsilon(n,\, S, \, w_{\mu} w^{\mu})/\partial S$;
and $\Lambda \equiv 2 \, \partial \varepsilon(n,\, S, \, w_{\mu} w^{\mu})/\partial (w_{\mu} w^{\mu})$.
Equation (\ref{2ndlaw}) is interpreted 
as the second law of thermodynamics for a superfluid liquid.

We need also to specify the pressure $P$.
According to the standard definition 
it equals to a partial derivative of the full system energy 
$\varepsilon V$ with respect to volume $V$ at constant 
total number of particles, total entropy, and $w_{\mu}w^{\mu}$ \cite{ab76, khalatnikov00},
\begin{equation}
P \equiv -\frac{\partial \left(\varepsilon V \right)}{\partial V} = -\varepsilon +\mu n + TS.
\label{pres}
\end{equation}
Using (\ref{2ndlaw}) and (\ref{pres})
one arrives at the following Gibbs-Duhem equation 
for a superfluid liquid,
\begin{equation}
dP = n  \, d\mu + S \, dT - \frac{\Lambda}{2} \, d \left( w_{\mu} w^{\mu} \right).
\label{dP}
\end{equation}
%

\subsection{Determination of $\Delta T^{\mu\nu}$ 
and the parameters $B$ and $\Lambda$}
\label{sec2b}

We discussed above a general structure of the {\it non-dissipative}
hydrodynamics of superfluid liquid,
which must conserve entropy of any closed system.
This means that the entropy generation equation must take the form
of the continuity equation,
\begin{equation}
\partial_{\mu} S^{\mu}=0,
\label{entropy}
\end{equation}
where $S^{\mu}$ is the entropy current density
(it will be shown below that the entropy flows with the normal liquid component,
i.e.\, $S^{\mu}=S u^{\mu}$). 

We will find $\Delta T^{\mu\nu}$, $B$, and $\Lambda$
from this requirement.
To do this, 
we should derive the entropy generation equation from the hydrodynamics 
of the previous section.
Let us consider a combination $u_{\nu} \partial_{\mu} T^{\mu\nu}$,
which vanishes in view of Eq.\ (\ref{Tmunu}). 
Using Eqs.\ (\ref{Tmunu2}), (\ref{norm1}), (\ref{pres}), and (\ref{dP})
one obtains
\begin{equation}
0= -u^{\mu} \, T\, \partial_{\mu} S - ST \, \partial_{\mu}u^{\mu}
- \mu \, \partial_{\mu}(n u^{\mu})-u^{\mu}\, \Lambda \, w_{\nu} \, \partial_{\mu} w^{\nu} 
+ u_{\nu} \, \partial_{\mu}\Delta T^{\mu\nu},
\label{entr1xc}
\end{equation}
or, using Eq.\ (\ref{jmu}) with $j^{\mu}$ from Eq.\ (\ref{jmu4}),
\begin{equation}
T \, \partial_{\mu} (S u^{\mu}) = \mu \, \partial_{\mu}(Y w^{\mu})
-u^{\mu} \, \Lambda \, w_{\nu} \, \partial_{\mu} w^{\nu}
+u_{\nu} \, \partial_{\mu}\Delta T^{\mu\nu}.
\label{entr2xc}
\end{equation}
This equation can be further 
transformed to
\begin{equation}
T \, \partial_{\mu} (S u^{\mu}) = 
\partial_{\mu}(\mu \, Y w^{\mu}) - Y w^{\mu}\, \partial_{\mu}\mu 
-u^{\mu} \, \Lambda \, w_{\nu} \, \partial_{\mu} w^{\nu}
+ \partial_{\mu}(u_{\nu}\Delta T^{\mu\nu}) 
- \Delta T^{\mu\nu}\partial_{\mu}u_{\nu}.
\label{entr3}
\end{equation}
The derivative $\partial_{\mu} w^{\nu}$
in the third term on the right-hand side
of Eq.\ (\ref{entr3}) can be expressed 
by making use of Eq.\ (\ref{weq}).
After substitution of the result a few terms 
vanish 
and we left with 
\begin{equation}
T \, \partial_{\mu} (S u^{\mu}) = 
w^{\mu}(\Lambda \, \partial_{\mu}B - Y \, \partial_{\mu}\mu)
+ \partial_{\mu} \left( \mu Y w^{\mu} + u_{\nu} \, \Delta T^{\mu\nu} \right)
+ \partial_{\mu} u_{\nu} \, \left( \Lambda \, w^{\mu} w^{\nu} 
+ \Lambda \, B \, u^{\mu} w^{\nu} -\Delta T^{\mu\nu} \right).
\label{entr4}
\end{equation}
To obtain Eq.\ (\ref{entr4}) we used the equalities
\begin{eqnarray}
u_{\mu} \, \partial_{\nu} u^{\mu} &=&0,
\label{11}\\
u_{\mu} \, \partial^{\nu} w^{\mu} &=& -w^{\mu} \, \partial^{\nu}u_{\mu},
\label{22}
\end{eqnarray}
following from Eqs.\ (\ref{norm1}) and (\ref{wu}), respectively.
The second and third terms in Eq.\ (\ref{entr4})
can be symmetrized by employing Eqs.\ (\ref{norm1}) and (\ref{wu}).
As a result, Eq.\ (\ref{entr4}) can be rewritten in its final form as
\begin{eqnarray}
T \, \partial_{\mu} (S u^{\mu}) &=& 
w^{\mu}(\Lambda \, \partial_{\mu}B - Y \, \partial_{\mu}\mu)
\nonumber\\
&+& \partial_{\mu}\left[ 
u_{\nu} \left(\Delta T^{\mu\nu} - \Lambda w^{\mu} w^{\nu} - \mu Y w^{\mu} u^{\nu} - \mu Y w^{\nu} u^{\mu}\right)
\right]
\nonumber\\
&+& \partial_{\mu} u_{\nu} \, \left( \Lambda \, w^{\mu} w^{\nu} 
+ \Lambda \, B \, u^{\mu} w^{\nu} + \Lambda \, B \, u^{\nu} w^{\mu} -\Delta T^{\mu\nu} \right)
\label{entr5}
\end{eqnarray}
or
\begin{eqnarray}
\partial_{\mu} (S u^{\mu}) &=& 
\frac{w^{\mu}}{T}(\Lambda \, \partial_{\mu}B - Y \, \partial_{\mu}\mu)
+ \left(\mu Y-\Lambda B \right) \, \frac{\partial_{\mu}T}{T^2} \, w^{\mu} 
\nonumber\\
&+& \partial_{\mu}\left[ 
\frac{u_{\nu}}{T} \left(\Delta T^{\mu\nu} - \Lambda w^{\mu} w^{\nu} - \mu Y w^{\mu} u^{\nu} - \mu Y w^{\nu} u^{\mu}\right)
\right]
\nonumber\\
&+& \partial_{\mu}\left( \frac{u_{\nu}}{T} \right) \, \left( \Lambda \, w^{\mu} w^{\nu} 
+ \Lambda \, B \, u^{\mu} w^{\nu} + \Lambda \, B \, u^{\nu} w^{\mu} -\Delta T^{\mu\nu} \right).
\label{entr5b}
\end{eqnarray}
The right-hand side of this equation must be a four-divergence
for {\it any} $\partial_\mu u_{\nu}$, $\partial_\mu T$, and $\partial_{\mu}\mu$.
This requirement, together with the assumption that $\Delta T^{\mu\nu}$ 
should depend on the four-velocities $u^{\mu}$ and $w^{\mu}$ 
and various thermodynamic quantities (but not on their gradients!),
while $B$ and $\Lambda$ should depend on thermodynamic quantities only,
allows us to identify the unknown parameters 
$\Lambda$, $B$, $\Delta T^{\mu\nu}$, and $S^{\mu}$ as
\begin{eqnarray}
\Lambda &=& \frac{Y}{k},
\label{Lam}\\
B &=& k \mu,
\label{BB}\\
\Delta T^{\mu\nu} &=& Y \left( \frac{w^{\mu} w^{\nu}}{k} + \mu u^{\mu} w^{\nu} + \mu u^{\nu} w^{\mu} \right),
\label{entr6}\\
S^{\mu} &=& S u^{\mu},
\label{Smu2}
\end{eqnarray}
where $k$ is some constant which should be equal to 1, 
as follows from the comparison with the non-relativistic theory.%
%
\footnote{
Another way to verify that $k$ can be chosen equal to 1 is to note
that {\it both} $\phi$ and $Y$ are introduced into the theory
through the definition (\ref{ji}) of $j^i$ in the comoving frame.
They can, therefore, be simultaneously rescaled,  
$Y \rightarrow Y/k$ and
$\phi \rightarrow k \phi$, 
without affecting $j^i$ and other observables of the theory.
This is equivalent to choosing $k=1$ in Eqs.~(\ref{Lam})--(\ref{entr6}).}
%
These equalities complete the formulation of relativistic superfluid hydrodynamics
in the absence of vortices.
One can see that the resulting energy-momentum tensor $T^{\mu\nu}$,
\begin{equation}
T^{\mu\nu} = (P+\varepsilon) u^{\mu} u^\nu + P g^{\mu\nu} + 
Y \left( w^{\mu} w^{\nu} + \mu u^{\mu} w^{\nu} + \mu u^{\nu} w^{\mu} \right),
\label{Tmunu3}
\end{equation}
is symmetric and satisfies the condition (\ref{condTmunu}) 
(on account of Eq.\ \ref{wu}).

Eqs.\ (\ref{2ndlaw}) and (\ref{dP}) now take the form 
\begin{eqnarray}
d \varepsilon &=& \mu \, dn + T \, dS + \frac{Y}{2} \,  d \left( w_{\mu} w^{\mu} \right),
\label{2ndlawXXX}\\
dP &=& n  \, d\mu + S \, dT - \frac{Y}{2} \, d \left( w_{\mu} w^{\mu} \right),
\label{dPXXX}
\end{eqnarray}
while the potentiality condition (\ref{weq}) becomes
\begin{equation}
\partial_{\mu}(w_{\nu}+ \mu u_{\nu})=\partial_{\nu}(w_{\mu}+\mu u_{\mu})
\,\,\, \Leftrightarrow \,\,\, 
m\left[\partial_{\mu}V_{({\rm s})\, \nu}-\partial_{\nu}V_{({\rm s})\, \mu} \right]=0.
\label{weq2}
\end{equation}
%

\vspace{0.2 cm}
\noindent
%
{\bf Remark 1.}
It is relatively straightforward to include dissipation into
this hydrodynamics. 
The corresponding corrections (the largest of them) 
have been first obtained in Refs.\ \cite{lk82, kl82}
and have received a great deal of attention in the recent years 
\cite{pd03,gusakov07,bbm11,hlsy11,bbmy14}.
For the superfluid hydrodynamics in the form discussed above 
they were formulated in Ref.\ \cite{gusakov07}.

Dissipation adds a correction $\tau^{\mu\nu}_{\rm diss}$ 
to the energy-momentum tensor $T^{\mu\nu}$ (\ref{Tmunu3})
and also changes the relation between the superfluid velocity $V^{\mu}_{({\rm s})}$
and the four-vector $w^{\mu}$, which becomes \cite{gusakov07}\,%
%
\footnote{In the absence of dissipation 
$V^{\mu}_{({\rm s})}=(w^{\mu}+\mu u^{\mu})/m$,
as follows from Eqs.\ (\ref{Vs}), (\ref{new}), and (\ref{BB}).} 
%
%
\begin{equation}
V^{\mu}_{({\rm s})}=\frac{w^{\mu}+(\mu+\varkappa_{\rm diss}) u^{\mu}}{m},
\label{sflr1}
\end{equation}
where $\varkappa_{\rm diss}$ is the correction depending on the bulk viscosity
coefficients $\xi_3$ and $\xi_4$.
Both these corrections are briefly discussed
in Appendix \ref{full_hydro}, 
where we 
present
the full system of equations 
of relativistic superfluid HVBK-hydrodynamics.

\section{Relativistic superfluid hydrodynamics 
in the presence of vortices}

A thorough discussion of vortices in the non-relativistic superfluid hydrodynamics 
can be found in many references (see, e.g., \cite{khalatnikov00,donnelly05,sonin87,varoquaux15});
a brief summary of results is given in Appendix \ref{HVBK}.
An extension of the concept of vortices to the relativistic case 
is rather straightforward (see, e.g., Refs.\ \cite{rothen81, lk82, kl82, cl95, carter00}).
When there are no vortices in the system the wave function phase of 
a superfluid condensate is a well-defined quantity everywhere so that
the integral
$\oint \partial_{\mu} \phi \, dx^{\mu}$
over {\it any} closed loop vanishes. 
If there are topological defects -- vortices -- in the system,
this integral should not be necessarily zero
and can be a multiple of $2 \pi$ 
(it cannot be arbitrary in order 
for the wave function of the condensate 
to be uniquely defined),%
\begin{equation}
\oint V_{({\rm s})\, \mu} \, dx^{\mu} =\frac{2 \pi N}{s m},
\label{int22}
\end{equation}
where $N$ is an integer; 
$s=1$ for Bose-superfluids and $s=2$ for Fermi-superfluids,
and we introduced the superfluid velocity $V_{({\rm s}) \, \mu}$
instead of $\partial_{\mu} \phi$ 
(sufficiently far from the vortices, 
where the ``hydrodynamic approach'' is justified, 
they are related by Eq.~\ref{Vs};
however, in the immediate vicinity of the vortex cores
this equation is violated \cite{ll80}). 
It can be shown \cite{khalatnikov00} that in a real superfluid it is energetically 
favorable to form vortices in the form of thin lines, 
each carrying exactly one quantum of circulation 
[i.e., an integral \ref{int22} over a closed loop around any given vortex line is $2 \pi/(s m)$].

Equation (\ref{int22}) can be rewritten, using the Stokes' theorem, 
as an integral over the surface encircled by the loop,
\begin{equation}
\int d f^{\mu \nu} 
F_{\mu \nu}
= \frac{2 \pi N}{s},
\label{int33}
\end{equation}
where $F_{\mu\nu}$ defines vorticity multiplied by $m$,
$F_{\mu\nu} \equiv m [\partial_{\mu} V_{({\rm s})\, \nu} -\partial_{\nu}V_{({\rm s})\,\mu}]$
(for brevity, $F_{\mu\nu}$ is called ``vorticity'' in what follows).
In many physically interesting situations%
\footnote{For example, in rotating neutron stars, 
the mean distance between the neighboring vortices is $\sim 10^{-2}-10^{-4}$~cm, while 
the typical length-scale, the stellar radius, is $\sim 10$~km.
} 
vortices are so densely packed on a typical length-scale of the problem
that it makes no sense to follow the evolution of each of them in order to describe
dynamics of the system as a whole.
Instead, it is more appropriate to 
use {\it coarse-grained}
dynamical equations
which depend on 
quantities averaged over 
the volume containing large amount of vortices.

The main parameters of such a theory are 
the smooth-averaged superfluid velocity 
and vorticity
(to be defined
as $V_{({\rm s})}^{\mu}$ and $F_{\mu\nu}$ in what follows);
they are analogous to, respectively, 
the averaged superfluid velocity ${\pmb V}_{\rm s}$ and $m\, {\rm curl} \,  {\pmb V}_{\rm s}$
of the nonrelativistic theory.
Note that, in view of Eq.\ (\ref{int33}), the smooth-averaged vorticity $F_{\mu\nu}\neq 0$ 
(and $V_{({\rm s})}^{\mu}$ is not simply given by a gradient of scalar). 
In other words, when there are vortices in the system, 
Eq.\ (\ref{weq2}) should be replaced by a weaker constraint (see below).

\subsection{Hydrodynamic equations 
under condition that 
the vortex contribution to the energy density
can be neglected}
\label{neglected}

To get an insight into the problem, 
let us first determine the form of 
large-scale hydrodynamics
in the case when one can neglect contribution of vortices 
to the second law of thermodynamics and 
to the energy-momentum tensor%
%
\footnote{For clarity, we also ignore in what follows 
the standard viscous and thermal conduction terms 
in the expression for $T^{\mu\nu}$ ($\tau^{\mu\nu}_{\rm diss}=0$)
and in the relation (\ref{sflr1}) between $V_{({\rm s})}^{\mu}$ and $w^{\mu}$ 
($\varkappa_{\rm diss}=0$).
}.
%
In the nonrelativistic theory this limit corresponds to HVBK-hydrodynamics 
with $\lambda=0$ (when $\hbar$ is formally set to 0; see Appendix \ref{HVBK} and Remark 2 there).
In this limit vortices 
affect 
only the superfluid equation (\ref{weq2}) 
(Eq.\ \ref{sfleq1} of the nonrelativistic theory), 
while other equations of Sec.~\ref{Sec2} remain 
unchanged.
Note, however, that now these equations 
depend on the smooth-averaged four-velocity $V_{({\rm s})}^{\mu}$ which
is not given by simply $\partial^{\mu} \phi/m$.
Correspondingly, the smooth-averaged quantity $w^{\mu}$ in these equations
should now be written as 
(see Eq.\ \ref{sflr1} with $\varkappa_{\rm diss}=0$)
\begin{equation}
w^{\mu} = m V_{({\rm s})}^{\mu}-\mu u^{\mu}.
\label{wmu2}
\end{equation}

To find an explicit form of the smooth-averaged superfluid equation 
in the presence of vortices 
we will again make use of the fact that the entropy of a closed system cannot decrease.
Employing the energy-momentum and particle conservation laws (\ref{Tmunu}) and (\ref{jmu})
with $j^{\mu}$ and $T^{\mu\nu}$ given by, respectively, 
Eqs.\ (\ref{jmu4}) and (\ref{Tmunu3}), 
as well as Eqs.\ (\ref{norm1}), (\ref{wu}), (\ref{pres}), 
(\ref{2ndlawXXX}), and (\ref{dPXXX}),
we arrive at the following entropy generation equation,%
\begin{equation}
T \partial_{\mu}(S u^\mu) = u^\nu \, Y w^\mu \, F_{\mu\nu}.
\label{entropy5}
\end{equation}
This equation can be derived in the same way 
as in Sec.\ \ref{sec2b} with the only difference 
that now it is obtained without making use 
of the potentiality condition (\ref{weq2}),
which is not valid in the system with the distributed vorticity 
($F_{\mu\nu} \neq 0$).

Because entropy does not decrease, one should have
\begin{equation}
u^\nu \, Y w^\mu F_{\mu\nu}\geq 0.
\label{entropy_enq1}
\end{equation}
Let us now introduce a new four-vector,
\begin{equation}
f_\mu \equiv \frac{u^\nu F_{\mu\nu}}{\mu n }. 
\label{f01}
\end{equation}
In terms of $f_{\mu}$
Eq.\ (\ref{entropy_enq1}) can be rewritten as 
\begin{equation}
W^\mu f_{\mu}\geq 0,
\label{entr2}
\end{equation}
where we also defined
\begin{equation}
W^{\mu} \equiv \frac{Y w^{\mu}}{n}.
\label{W1}
\end{equation}
In the comoving frame 
[where $u^\mu=(1,0,0,0)$], 
$f_0=F_{00}=0$ 
and Eq.\ (\ref{entr2}) transforms into 
\begin{equation}
W^i f_i\geq 0,
\label{ineqX1}
\end{equation}
where $i=1,\, 2,\, 3$ is the spatial index.%
%
\footnote{It is worth noting that, 
in view of Eq.\ (\ref{wu}), 
$W^0=Y  w^0/n$ also vanishes 
in the comoving frame, $W^0=0$.}
%
In order for the inequality (\ref{ineqX1}) 
to hold true
the vector ${\pmb f} \equiv (f^1,\, f^2,\, f^3)$ 
should satisfy a number of conditions
(forget for a moment about its definition~\ref{f01}):
($i$) it must be polar; 
($ii$) must vanish at $F_{\mu\nu}=0$ (because the potentiality condition \ref{weq2}
is valid in that case);
and 
($iii$) should depend on ${\pmb W} \equiv (W^1,\, W^2,\, W^3)$ 
in order to satisfy 
Eq.\ (\ref{ineqX1})
at arbitrary ${\pmb W}$ 
(note that ${\pmb V}_{\rm s} \equiv 
[V_{(\rm s)}^1,\, V_{(\rm s)}^2, \, V_{(\rm s)}^3] =n {\pmb W}/(Y m)$ 
and thus is not an independent variable; see Eqs.\ \ref{wmu2} and \ref{W1}).
These conditions are clearly insufficient to determine the 
most general form of ${\pmb f}$.
However, 
it seems reasonable to further require that 
($iv$) ${\pmb f}$
may only depend on ${\pmb W}$ and $F_{\mu\nu}$ 
(as noted by Clark \cite{clark63}, 
in the non-relativistic theory
a similar assumption was implicitly made in Ref.\ \cite{bk61}; 
see Ref.\ \cite{clark63} for a detailed critical analysis
of HVBK-hydrodynamics).

In analogy with electrodynamics, instead of the antisymmetric tensor 
$F_{\mu\nu} = m [\partial_{\mu} V_{({\rm s})\, \nu} -\partial_{\nu}V_{({\rm s})\,\mu}]$
it is convenient to introduce an axial vector 
${\pmb H}=m \, {\rm curl} \,{\pmb V}_{\rm s}$
and a polar vector 
${\pmb E} \equiv m \, \left[\partial {\pmb V}_{\rm s}/\partial t + {\pmb \nabla} V_{({\rm s})}^0 \right]$.
Then the most general form of ${\pmb f}$, 
satisfying the conditions ($i$)--($iv$), 
can, in principle, be found. 
The resulting expression 
will contain many more kinetic coefficients (and additional terms)
in comparison to the original HVBK-expression (\ref{F}),
because now we allow ${\pmb f}$ to depend not only on ${\pmb H}=m \,  {\rm curl}\, {\pmb V}_{\rm s}$, 
like in the nonrelativistic theory, 
but also on the vector ${\pmb E}$
%
\footnote{
\label{EHterms}
Among the ${\pmb E}$-dependent terms which can enter the expression for ${\pmb f}$ 
there should be terms of the form 
$[{\pmb E}\times {\pmb W}]\times {\pmb W}$, $({\pmb E}{\pmb W}) \, {\pmb E}$,
$[{\pmb E}\times {\pmb W}]\times {\pmb E}$ and a number of ``mixed'' terms depending on both
${\pmb H}=m\,{\rm curl}\, {\pmb V}_{\rm s}$ and ${\pmb E}$, e.g.,
$[{\pmb W}\times{\pmb E}]\, ({\pmb E} {\pmb H})$ 
and ${\pmb W}\times[{\pmb E}\times {\pmb H}]\,({\pmb E} {\pmb H})$.
}.
%
The physical meaning of these additional terms is not clear and deserves a further study.
However, in the non-relativistic limit
these terms are presumably suppressed
in comparison to the ${\pmb H}$-dependent terms presented in Eq.\ (\ref{fifi}) below
(because ${\pmb E}\sim 1/c \rightarrow 0$ at $c\rightarrow \infty$, 
see Appendix \ref{nonrel_limit} and Eqs.\ \ref{orthNR} and \ref{spatialNR4} there).
Since here we are mainly interested in the straightforward generalization of HVBK-equations
to the relativistic case, below we only present the terms which have 
direct counterparts in the nonrelativistic theory.
They exclusively depend on the vector ${\pmb H}=m \, {\rm curl} \,{\pmb V}_{\rm s}$, 
namely,
\begin{eqnarray}
{\pmb f}=-\alpha\, [{\pmb H} \times {\pmb W}]
-\beta\,  {\pmb {\rm e}} \times[{\pmb H} \times {\pmb W}]+
\gamma\,   {\pmb {\rm e}} ({\pmb W}\,{\pmb H}),
\label{fifi}
\end{eqnarray}
where 
${\pmb {\rm e}}  \equiv {\pmb H}/H$
is the unit vector
in the direction of ${\pmb H}=m \, {\rm curl} \, {\pmb V}_{\rm s}$;
$\alpha$, $\beta$, and $\gamma$ are some scalars 
(kinetic coefficients),
which can generally depend on invariants of
${\pmb W}$ and ${\pmb H}$. 
Note that the first term in the right-hand side of Eq.\ (\ref{fifi}), 
depending on $\alpha$, is dissipationless.
In contrast, the other terms there 
are dissipative
and to satisfy (\ref{entr2}) the coefficients 
$\beta$ and $\gamma$ should be positive, 
$\beta,\gamma \geq 0$.
In appendix \ref{nonrel_limit} it is shown 
that these coefficients indeed coincide with the 
coefficients $\alpha$, $\beta$, and $\gamma$ of HVBK-hydrodynamics.

We found the form of the four-vector $f^{\mu}$ in the comoving frame, 
$f^{\mu}=(0, \, {\pmb f})$, where ${\pmb f}$ is given by Eq.\ (\ref{fifi}).
Now our aim will be to rewrite $f^{\mu}$ in an arbitrary frame.
To do this let us introduce a four-vector $H^{\mu}$, 
given by (in the orthonormal basis)
\begin{equation}
H^{\mu} \equiv \epsilon^{\mu \nu \lambda \eta} \, u_{\nu} \, 
m \, \partial_{\lambda} V_{(\rm s) \eta}
= \frac{1}{2} \, \epsilon^{\mu \nu \lambda \eta} \, u_{\nu} \, F_{\lambda \eta},
\label{Hmu}
\end{equation}
where $\epsilon^{\mu \nu \lambda \eta}$ is the four-dimensional Levi-Civita tensor
and we use the anti-symmetry property of the tensor $F_{\mu\nu}$
in the second equality.
In the comoving frame this vector equals 
$H^{\mu}=(0,\, {\pmb H})=(0, \, m \, {\rm curl}\,  {\pmb V}_{\rm s})$.
Also, assume that we have two four-vectors, say, $B^{\mu}$ and $C^{\mu}$, 
whose spatial components ${\pmb B}$ and ${\pmb C}$ 
form a 3D-vector ${\pmb A}={\pmb B}\times {\pmb C}$ in the comoving frame.
Then we {\it define} the four-vector $A^{\mu}$ in an arbitrary frame according to
\begin{equation}
A^{\mu} \equiv \epsilon^{\mu \nu \lambda \eta} \, u_{\nu} \, B_{\lambda} \, C_{\eta}.
\label{A}
\end{equation}
The definitions (\ref{Hmu}) and (\ref{A}) are trivial extensions
of the curl operator and cross product, 
defined in the comoving frame [$u^{\mu}=(1,0,0,0)$], 
to an arbitrary frame 
(see also Refs.\ \cite{lichnerowicz67, ehlers93} for similar definitions).
Using these definitions, one can immediately write out a 
general
Lorentz-covariant expression for $f^{\mu}$,
\begin{equation}
f^{\mu} = -\alpha \, X^{\mu}
- \beta \, \epsilon^{\mu\nu\lambda\eta} \, u_{\nu} \, {\rm e}_{\lambda} \, X_{\eta}
+ \gamma \, {\rm e}^{\mu} \, (W^{\lambda} H_{\lambda}),
\label{finv}
\end{equation}
where ${\rm e}^{\mu}=H^{\mu}/H$ with $H=(H_{\mu} H^{\mu})^{1/2}$
and $X^{\mu} \equiv \epsilon^{\mu\nu\lambda\eta} \, u_{\nu} \, H_{\lambda} \, W_{\eta}$.%
\footnote{Possible ${\pmb E}$-dependent terms in the expression for $f^{\mu}$ 
(see the footnote \ref{EHterms}) can be obtained in a similar way 
by introducing a four-vector $E^{\nu}\equiv u_{\mu}F^{\mu\nu}$, 
which reduces to $(0,{\pmb E})$ in the comoving frame.}
The same expression can be reformulated
without making use of the Levi-Civita tensor,%
%
\footnote{Eq.\ (\ref{f2}) is the most general expression for $f^{\mu}$ valid
for arbitrary $W^{\mu}$.
However, the four-vector $W^{\mu}$, 
introduced in this section 
(cf.\ the definition of $W^{\mu}$ in Sec.\ \ref{included}), 
satisfies a condition $u_{\mu} W^{\mu}=0$ (see Eqs.\ \ref{wu} and \ref{W1}),
which allows one to simplify Eq.\ (\ref{f2}) in this particular case 
and write
\begin{equation}
f^{\mu} = \alpha\perp^{\mu\nu} F_{\nu\lambda} W^\lambda
+ \frac{\beta-\gamma}{H}  \perp^{\mu\eta} \perp^{\nu\sigma}  F_{\eta\sigma} F_{\lambda\nu}  W^{\lambda}
+\gamma  H \, W^{\mu}.
\nonumber
\end{equation}
}
%
\begin{equation}
f^{\mu} = \alpha\perp^{\mu\nu} F_{\nu\lambda} \, W_{\delta} \perp^{\lambda \delta}
+ \frac{\beta-\gamma}{H}  \perp^{\mu\eta} \perp^{\nu\sigma}  F_{\eta\sigma} F_{\lambda\nu}  \,
W_{\delta} \perp^{\lambda \delta}
+\gamma  H \, W_{\delta} \perp^{\mu \delta},
\label{f2}
\end{equation}
where
$\perp^{\mu\nu}=g^{\mu\nu}+u^{\mu}u^{\nu}$ is the projection operator and
\begin{equation}
H=\sqrt{\frac{1}{2} \perp^{\mu\eta} \perp^{\nu\sigma} F_{\mu\nu}  F_{\eta\sigma}}.
\label{H2}
\end{equation}

Because $f^{\mu}$ is now specified, 
Eq.\ (\ref{f01}) can now be treated as 
a {\it new superfluid equation} 
which replaces the potentiality condition (\ref{weq2}) 
and generalizes it to the case of a superfluid liquid 
with distributed vorticity.
It can be rewritten as
\begin{equation}
u^\nu F_{\mu\nu}= \mu n \, f_\mu.
\label{sflrot1}
\end{equation}
Note that it is valid as long as one can neglect the contribution
of vortices to the energy density 
(i.e., $\lambda=0$, see Appendix \ref{evortex}).
Otherwise, the definition of the vector $W^{\mu}$ 
should be modified 
(see Eq.\ \ref{Wmunew} in Sec.\ \ref{included}).
In Appendix \ref{nonrel_limit} we demonstrate
that, in the nonrelativistic limit, 
Eq.\ (\ref{sflrot1}) reduces to Eq.\ (\ref{sfleq1}) with $\lambda=0$.

\vspace{0.2 cm}
\noindent
%
{\bf Remark 1.}
To derive the superfluid equation (\ref{sflrot1})
we first introduced the vector $f^{\mu}=u^{\nu}F_{\mu\nu}/(\mu n)$
and then deduced its possible form from the condition $f_{\mu}W^{\mu} \geq 0$.
This is not the only way of obtaining this equation.
In fact, Eq.\ (\ref{sflrot1}) can also be derived by introducing a vector
$g_{\nu}\equiv W^{\mu}F_{\mu\nu}$ and then 
requiring
it to satisfy a condition $g_{\nu}u^{\nu}\geq 0$,
which follows from the constraint (\ref{entropy_enq1}).

\vspace{0.2 cm}
\noindent
%
{\bf Remark 2.}
Equation (\ref{sflrot1}) imposes 
certain restrictions on the possible form 
of the tensor $F_{\mu\nu}$. 
Assume 
that 
$F_{\mu\nu}$ satisfies this equation.
Then it can be shown by direct calculation 
that, if the coefficient $\gamma$ in Eq.\ (\ref{f2}) vanishes,
then 
a four-vector $V_{({\rm L})}^{\mu}$ exists, 
given by,
\begin{equation}
V_{({\rm L})}^{\mu}= u^{\mu} - \mu n \, \alpha \, W_{\nu} \perp^{\mu\nu}
+\frac{\mu n \, \beta}{H} \, \perp^{\mu \alpha} \perp^{\nu \beta} \, F_{\alpha\beta} \, W_{\nu},
\label{VlX0}
\end{equation}
such that the combination
$V_{({\rm L})}^{\nu} F_{\mu\nu}$ is identically zero,
\begin{equation}
V_{({\rm L})}^{\nu} F_{\mu\nu}=0.
\label{VlX1}
\end{equation}
Equation (\ref{VlX1}) is analogous to the vorticity conservation equation (\ref{rotomega})
of the non-relativistic HVBK-hydrodynamics (see Appendix \ref{HVBK}).

\vspace{0.2 cm}
\noindent
%
{\bf Remark 3.}
In Appendix \ref{HVBK} we consider the strong and weak-drag limits for
superfluid equation (\ref{sfleq1}) (or \ref{rotomega}) 
of the non-relativistic HVBK-hydrodynamics.
Similar limits can also be considered in relativistic hydrodynamics.
In particular, strong-drag limit corresponds to $\alpha=\beta=\gamma=0$ in Eq.\ (\ref{f2})
so that 
Eq.\ (\ref{VlX1}) 
reduces to
\begin{equation}
u^{\nu}F_{\mu\nu}=0.
\label{strong1}
\end{equation}
This equation describes vortex motion (vorticity transfer)
with the velocity $u^{\mu}$ of normal liquid component.
(In ordinary nonsuperfluid hydrodynamics a similar equation takes place, 
but vorticity $F_{\mu\nu}$ there is expressed through the same velocity $u^{\mu}$,
with which it is transferred,
$F_{\mu\nu}=\partial_{\mu}(\mu u_{\nu})-\partial_{\nu}(\mu u_{\mu})$, 
see, e.g., Ref.\ \cite{ac07}.)
Weak-drag limit is described by the equation
\begin{equation}
V_{({\rm s})}^{\nu}F_{\mu\nu}=0
\label{weak1}
\end{equation}
and follows from 
Eq.\ (\ref{VlX1})
when $\alpha=-1/(\mu^2 Y)$ and $\beta=\gamma=0$ 
(cf.\ the corresponding limit in the non-relativistic HVBK-hydrodynamics).
It corresponds to a vortex motion with the superfluid velocity $V_{({\rm s})}^{\mu}$.
Note that both these limits were analyzed in Ref.\ \cite{lsc98}
in application to zero-temperature superfluid neutron stars%
%
\footnote{In the ``weak-drag'' equation (33) of Ref.\ \cite{lsc98} one finds the 
total neutron current density instead of the superfluid velocity $V_{({\rm s})}^{\mu}$.
This is not surprising since the authors of Ref.\ \cite{lsc98} work in the limit $T=0$,
when all particles (neutrons) are paired
and move with one and the same superfluid velocity $V_{({\rm s})}^{\mu}$.
}.
%

\subsection{Accounting for the vortex energy}
\label{included}

In this section we formulate the relativistic generalization
of the HVBK-hydrodynamics taking into account contribution
of vortices to the energy-density $\varepsilon$ 
and the energy-momentum tensor $T^{\mu\nu}$.
It is convenient to formulate this hydrodynamics 
in terms of the four-vectors $u^{\mu}$ and $w^{\mu}$ 
as primary degrees of freedom.
For that it is necessary to define more rigorously
what we actually mean by $w^{\mu}$.
In what follows we {\it define} $w^{\mu}$ by the formula 
\begin{equation}
j^{\mu}=n u^{\mu} + Y w^{\mu},
\label{jmu22}
\end{equation}
where the quantities $j^{\mu}$, $u^{\mu}$, $n$
have the same meaning as in the previous sections 
[in particular, $n$ is the number
density measured in the comoving frame where $u^{\mu}=(1,0,0,0)$],
while the parameter $Y$ 
is defined by the second law of thermodynamics (see Eq.\ \ref{2ndlaw2} below),
$Y=2 \partial \varepsilon/\partial(w_{\mu} w^{\mu})$.
(It is straightforward to show that it is always possible 
to define $w^{\mu}$ by Eq.\ \ref{jmu22}
such that the coefficients $Y$ in Eq.\ \ref{jmu22} and $Y$ in Eq.\ \ref{2ndlaw2} 
will indeed coincide.)
A definition (\ref{jmu22}) implies that the conditions
(\ref{condjmu2}) and (\ref{wu}) must be satisfied automatically.

After defining $w^{\mu}$, 
the superfluid velocity $V_{({\rm s})}^{\mu}$ 
of our smooth-averaged hydrodynamics can be defined by Eq.\ (\ref{wmu2}),
$V_{({\rm s})}^{\mu}=(w^{\mu}+\mu u^{\mu})/m$.%
%
\footnote{
This way of reasoning is similar 
to that of Bekarevich \& Khalatnikov \cite{bk61}.
In a purely phenomenological approach 
it is not obvious, however, that the superfluid velocity 
$V_{({\rm s})}^{\mu}$ defined in this manner
will coincide with the velocity,
whose vorticity is directly related to the area density of vortex lines
and satisfies, for example, the ``continuity equation'' for vortices (see Eq.\ \ref{VlX1}).
The fact that both definitions coincide follows from the self-consistency 
of the resulting hydrodynamics 
(in particular, Eq.\ \ref{VlX1} remains to be satisfied, see below).
This conclusion can also be verified by a microscopic consideration 
similar to that presented in Appendices \ref{evortex} and \ref{micro}.
}
%
(We again ignore here a viscous dissipative correction 
$\varkappa_{\rm diss}$, 
which has the same form \cite{gusakov07} 
as in the vortex-free case and
does not affect our derivation; 
it can easily be included in the final equations, 
see Appendix \ref{full_hydro}.)

Next, we present the energy-momentum tensor in the form 
\begin{equation}
T^{\mu\nu} = (P+\varepsilon) u^{\mu} u^\nu + P g^{\mu\nu} + 
Y \left( w^{\mu} w^{\nu} + \mu u^{\mu} w^{\nu} + \mu u^{\nu} w^{\mu} \right) 
+ \tau^{\mu\nu},
\label{Tmunu4}
\end{equation}
where $P$ is defined by Eq.\ (\ref{pres}) and 
$\tau^{\mu\nu}(=\tau^{\nu\mu})$ 
is the symmetric vortex contribution to $T^{\mu\nu}$, 
which will be determined below
(without this contribution Eq.\ \ref{Tmunu4} coincides with \ref{Tmunu3}).
Because $\varepsilon$ is the total energy density 
in the comoving frame
(including the contribution of vortices),
$T^{\mu\nu}$ should satisfy condition (\ref{condTmunu})
which, in view of Eq.\ (\ref{wu}), translates into
\begin{equation}
u_{\mu}u_{\nu}\tau^{\mu\nu}=0.
\label{taucond1}
\end{equation}

Finally, the most important step in building up
the relativistic HVBK-hydrodynamics is to 
postulate the form of the second law of thermodynamics
in the presence of vortices.
Obviously, one can write
\begin{equation}
d \varepsilon = \mu \, dn + T \, dS + \frac{Y}{2} \,  d \left( w_{\mu} w^{\mu} \right) 
+ d\varepsilon_{\rm vortex},
\label{2ndlaw2}
\end{equation}
where $d\varepsilon_{\rm vortex}$ is the term responsible 
for the vortex contribution to $d\varepsilon$,
while other terms are the same as in the vortex-free superfluid hydrodynamics
(see Eq.\ \ref{2ndlawXXX}).

Before guessing a possible form of $d\varepsilon_{\rm vortex}$
let us derive the entropy generation equation. 
Using equations of this section together with
Eqs.\ (\ref{Tmunu}), (\ref{jmu}), (\ref{norm1}), (\ref{wu}), and (\ref{pres}),
one gets
\begin{equation}
T \, \partial_{\mu} (S u^{\mu}) = 
u^{\nu}\, Y w^{\mu} \, F_{\mu\nu}
- u^{\mu} \, \partial_\mu \varepsilon_{\rm vortex} + u_{\nu} \, \partial_{\mu} \tau^{\mu\nu},
\label{entropy6}
\end{equation}
where%
%
\footnote{If $\varkappa_{\rm diss}$ were non-zero, 
one would have a combination
$F_{\mu\nu}-\partial_{\mu}(\varkappa_{\rm diss}u_{\nu})
+\partial_{\nu}(\varkappa_{\rm diss}u_{\mu})$
instead of $F_{\mu\nu}$ in Eq.\ (\ref{entropy6}).
}
%
%
\begin{equation}
F_{\mu\nu} \equiv m [\partial_{\mu} V_{({\rm s})\, \nu} -\partial_{\nu}V_{({\rm s})\,\mu}]
=\partial_{\mu}(w_{\nu}+\mu u_{\nu})
-\partial_{\nu}(w_{\mu}+\mu u_{\mu}).
\label{Fmunu3}
\end{equation}
The first term in the right-hand side of Eq.\ (\ref{entropy6})
is the same as in Eq.\ (\ref{entropy5}), 
the second and third terms are induced by the 
vortex-related terms in Eqs.\ (\ref{Tmunu4}) and (\ref{2ndlaw2}). 

Now let us specify $d \varepsilon_{\rm vortex}$.
In the absence of vortices the energy density $\varepsilon$ depends
on three scalars, $n$, $S$, and $w_{\mu}w^{\mu}$.
When vortices are present a new dynamical quantity 
$F_{\mu\nu}\neq0$ appears and the (smooth-averaged) 
energy density $\varepsilon$ 
can depend on its various invariants.
In fact, it is possible to compose many different scalars from the quantities
$F_{\mu\nu}$, $u^{\mu}$, $w^{\mu}$, and their derivatives.
One can single out one or few of them
on the basis of physical arguments or intuition.
As it is argued in Appendix \ref{evortex}, 
it is a good approximation 
to treat $\varepsilon_{\rm vortex}$ as a function of only one additional invariant 
$H=(H_{\mu}H^{\mu})^{1/2}$,
where $H^{\mu}$ is given by Eq.\ (\ref{Hmu})
and equals $(0, \, m\, {\rm curl} {\pmb V}_{\rm s})$
in the comoving frame.
Correspondingly, $H=m\, |{\rm curl} \, {\pmb V}_{\rm s}|$ 
is analogous to the invariant $\omega=|{\rm curl} \, {\pmb V}_{\rm s}|$
of the nonrelativistic theory (see Appendix \ref{HVBK}).
If $\varepsilon$ depends on $H$,
one can write
\begin{equation}
d \varepsilon_{\rm vortex} = \frac{\partial \varepsilon}{\partial H} \, d H
= \frac{\lambda}{2mH} d(H_{\mu}H^{\mu}),
\label{dEvort}
\end{equation}
where the partial derivative is taken at constant $n$, $S$, and $w_{\mu} w^{\mu}$;
$\lambda \equiv m \, \partial \varepsilon/\partial H$ is the relativistic analogue 
of the parameter $\lambda$ of the nonrelativistic theory 
(see Appendix \ref{evortex});
both parameters coincide in the nonrelativistic limit.

Eq.\ (\ref{dEvort}) can be rewritten as 
\begin{equation}
d \varepsilon_{\rm vortex} = 
\frac{\Gamma}{2} \,
\left( 
O^{\alpha\beta}\,d F_{\alpha\beta}
+ 2 \, F_{\alpha\beta} F^{\alpha \gamma} u^{\beta}  d u_{\gamma} \right),
\label{dEvort2}
\end{equation}
where we used Eq.\ (\ref{H2})
together with the identity $u_{\mu}u_{\nu}F^{\mu\nu}=0$, and defined%
%
\footnote{\label{footnote8}Note that 
$O^{\alpha\beta}$ can also be presented in the form, 
$O^{\alpha\beta}=\frac{1}{2} \, \epsilon^{\delta\eta\alpha\beta} \, u_{\eta} \,
\epsilon_{\delta a b c} \, u^a \, F^{bc}$ ($a$, $b$, and $c$ are the space-time indices).}
%
%
\begin{eqnarray}
\Gamma &\equiv& \frac{\lambda}{mH},
\label{Gamma2}\\
O^{\alpha\beta} &\equiv& \bot^{\alpha \gamma}\bot^{\beta	 \delta} 
\, F_{\gamma\delta}.
\label{Omunu}
\end{eqnarray}
In what follows we will be interested in the quantity
$-u^{\mu} \, \partial_{\mu} \varepsilon_{\rm vortex}$,
which appears in the entropy generation equation (\ref{entropy6}).
Using Eq.\ (\ref{dEvort2}), it is given by
\begin{equation}
-u^{\mu} \, \partial_{\mu} \varepsilon_{\rm vortex}
=- \frac{\Gamma}{2} \, u^{\mu} \, O^{\alpha\beta}\,\partial_{\mu} F_{\alpha\beta}
 -  \Gamma \, u^{\mu} u^{\delta}\, F_{\alpha\delta} F^{\alpha \nu}  \, \partial_{\mu} u_{\nu}.
\label{dedede}
\end{equation}
The first term in the right-hand side 
of Eq.\ (\ref{dedede}) can be transformed as
\begin{eqnarray}
-\frac{\Gamma}{2}\, u^{\mu} \, O^{\alpha \beta} \partial_\mu F_{\alpha \beta}
&=& u^{\nu} F_{\mu\nu} \, \, \partial_{\alpha}(\Gamma\, O^{\mu\alpha})
\nonumber\\
&-&\partial_{\mu}\left( u^{\nu} \, \Gamma \,  O^{\mu\alpha} F_{\nu\alpha} \right)
\nonumber\\
&+&\partial_{\mu}u^{\nu} \left( \Gamma \, O^{\mu\alpha} F_{\nu \alpha}
 \right).
\label{1term}
\end{eqnarray}
To obtain this expression we used the 
identity (see Eq.\ \ref{Fmunu3})
\begin{equation}
\partial_\mu F_{\alpha \beta} = 
\partial_{\alpha} F_{\mu \beta}+\partial_{\beta} F_{\alpha \mu}
\,\, \Leftrightarrow \,\, \epsilon^{iklm} \, \partial_k F_{lm}=0,
\label{equality}
\end{equation}
and the fact that 
both tensors $F^{\mu \nu}$ and $O^{\mu\nu}$ are antisymmetric.

In turn, the second term in the right-hand side of Eq.\ (\ref{dedede})
can be rewritten as
\begin{eqnarray}
-  \Gamma \, u^{\mu} u^{\delta}\, F_{\alpha\delta} F^{\alpha \nu} \, \partial_{\mu} u_{\nu} 
&=& 
- \Gamma \, \left[u^{\mu}u^{\delta}\, F_{\alpha \delta}F^{\alpha \nu} 
+ \underline{u^{\mu} u^{\nu} u^{\beta} u_{\gamma} \, F_{\alpha \beta} F^{\alpha \gamma} } \right]
 \partial_{\mu} u_\nu
\nonumber\\
&=& - \Gamma \,u^{\mu}u^{\gamma}
 \perp_{\nu\beta} F^{\alpha\beta} F_{\alpha\gamma} 
\,\,\partial_{\mu} u^\nu
\nonumber\\
&=& - \Gamma \,u^{\mu}u^{\gamma}
 \perp_{\nu\beta} F^{\alpha\beta} F_{\alpha\gamma} 
\,\,\partial_{\mu} u^\nu
+ \underline{\partial_{\mu} \left(\Gamma\, u^{\nu}
u^{\mu} u^{\gamma} \perp_{\nu\beta}  F^{\alpha\beta} F_{\alpha\gamma} \right)},
\label{2term}
\end{eqnarray}
where the underlined terms equal zero 
(because of Eq.\ \ref{11} and the equality $u^{\nu} \perp_{\nu\beta}=0$) 
and are added here 
in order to symmetrize the tensor $\tau^{\mu\nu}$
and to satisfy the condition (\ref{taucond1}), see below.
Using Eqs.\ (\ref{1term}) and (\ref{2term}), one obtains
\begin{eqnarray}
-u^{\mu} \, \partial_{\mu} \varepsilon_{\rm vortex} &=&
 u^{\nu} F_{\mu\nu} \, \, \partial_{\alpha}(\Gamma\, O^{\mu\alpha})
\nonumber\\
&-&\partial_{\mu}\left[ u^{\nu} \left(
\Gamma \,  O^{\mu\alpha} F_{\nu\alpha} 
- \Gamma \,u^{\mu}u^{\gamma}
\perp_{\nu\beta} F^{\alpha\beta} F_{\alpha\gamma} 
\right)\right]
\nonumber\\
&+&\partial_{\mu}u^{\nu} \left( \Gamma \, O^{\mu\alpha} F_{\nu \alpha}
- \Gamma \,u^{\mu}u^{\gamma}
\perp_{\nu\beta} F^{\alpha\beta} F_{\alpha\gamma} 
\right)
\nonumber\\
&=& u^{\nu} F_{\mu\nu} \, \, \partial_{\alpha}(\Gamma\, \bot^{\mu \gamma}\bot^{\alpha	 \delta} 
\, F_{\gamma\delta})
\nonumber\\
&-&\partial_{\mu}\left[ u_{\nu} \left(
 \Gamma \,  \perp_{\delta \alpha} F^{\mu\delta} F^{\nu \alpha}
-\Gamma \,  u^{\mu} u^{\nu} u^{\gamma} u_{\beta} F^{\alpha \beta} F_{\alpha \gamma}\right)\right]
\nonumber\\
&+&\partial_{\mu}u_{\nu} \left(  \Gamma \,  \perp_{\delta \alpha} F^{\mu\delta} F^{\nu \alpha}
-\Gamma \,  u^{\mu} u^{\nu} u^{\gamma} u_{\beta} F^{\alpha \beta} F_{\alpha \gamma}
\right),
\label{dex3}
\end{eqnarray}
where in the second equality 
we make use of the definition (\ref{Omunu}) for $O^{\alpha\beta}$.
Returning now to the entropy generation equation (\ref{entropy6}),
one can present it in the form
\begin{eqnarray}
T \, \partial_{\mu} (S u^{\mu}) &=& 
u^{\nu}F_{\mu\nu} \, \left[Y w^{\mu} 
+ \partial_{\alpha}(\Gamma\, \bot^{\mu \gamma}\bot^{\alpha	 \delta} 
\, F_{\gamma\delta})\right] 
\nonumber\\
&-& \partial_{\mu}\left[u_{\nu} \, (\Gamma\,   \perp_{\delta \alpha} F^{\mu\delta} F^{\nu \alpha}
- \Gamma \, u^{\mu} u^{\nu} u^{\gamma} u_{\beta} \, F^{\alpha \beta} F_{\alpha \gamma}
- \tau^{\mu\nu})
\right]
\nonumber\\
&+& \partial_{\mu}u_{\nu}  \left(
\Gamma\,   \perp_{\delta \alpha} F^{\mu\delta} F^{\nu \alpha}
- \Gamma \, u^{\mu} u^{\nu} u^{\gamma} u_{\beta} \, F^{\alpha \beta} F_{\alpha \gamma}
- \tau^{\mu\nu} \right)
\label{entropy2}
\end{eqnarray}
or
\begin{eqnarray}
\partial_{\mu} (S u^{\mu}) &=& 
\frac{u^{\nu}F_{\mu\nu}}{T} \, \left[Y w^{\mu} 
+ \partial_{\alpha}(\Gamma\, \bot^{\mu \gamma}\bot^{\alpha	 \delta} 
\, F_{\gamma\delta})\right] 
\nonumber\\
&-& \partial_{\mu}\left[\frac{u_{\nu}}{T} \, (\Gamma\,   \perp_{\delta \alpha} F^{\mu\delta} F^{\nu \alpha}
- \Gamma \, u^{\mu} u^{\nu} u^{\gamma} u_{\beta} \, F^{\alpha \beta} F_{\alpha \gamma}
- \tau^{\mu\nu})
\right]
\nonumber\\
&+& \partial_{\mu}\left(\frac{u_{\nu}}{T}\right)  \left(
\Gamma\,   \perp_{\delta \alpha} F^{\mu\delta} F^{\nu \alpha}
- \Gamma \, u^{\mu} u^{\nu} u^{\gamma} u_{\beta} \, F^{\alpha \beta} F_{\alpha \gamma}
- \tau^{\mu\nu} \right).
\label{entropy2b}
\end{eqnarray}
Neglecting dissipation, the right-hand side of this equation
should be a four-divergence
at arbitrary $\partial_{\mu} u_{\nu}$, $\partial_{\mu} T$, $w^{\mu}$, $\Gamma$, etc.
 This allows us to find%
%
\footnote{Unfortunately, 
detailed analysis shows that Eqs.\ (\ref{entr1}) and (\ref{tau_vortex}) 
do not follow unambiguously from Eq.\ (\ref{entropy2b}).
To obtain them unambiguously one needs to require, in addition,
that the spatial components $\tau^{ij}$ of the tensor $\tau^{\mu\nu}$
are
independent of the components $F_{0i}$ of the vorticity tensor 
{\it in the comoving frame}
(i.e., $\tau^{ij}$ there
depend on ${\pmb H}=m \, {\rm curl \,{\pmb V}_{\rm s}}$ only).
This additional assumption is confirmed 
by the results of independent microscopic consideration 
(see Appendix \ref{micro} and Remark 1 below in this section).
}
%
\begin{eqnarray}
&&u^{\nu}F_{\mu\nu} \, \left[Y w^{\mu} 
+ \partial_{\alpha}(\Gamma\, \bot^{\mu \gamma}\bot^{\alpha	 \delta} 
\, F_{\gamma\delta})\right] = 0 \quad {\rm and}
\label{entr1}\\
&&\tau^{\mu\nu} = \tau^{\mu\nu}_{\rm vortex}
=\Gamma\,   \perp_{\delta \alpha} F^{\mu\delta} F^{\nu \alpha}
- \Gamma \, u^{\mu} u^{\nu} u^{\gamma} u_{\beta} \, F^{\alpha \beta} F_{\alpha \gamma}.
\label{tau_vortex}
\end{eqnarray}
The first of these equations is similar to a non-dissipative version,
$u^\nu \, Y w^\mu F_{\mu\nu}= 0$,
of the condition (\ref{entropy_enq1}), 
analyzed in the previous section. 
It will clearly give us 
a (non-dissipative) superfluid equation generalized to the case when the terms depending on 
$\Gamma=\lambda/(mH)$ 
(see Eq.\ \ref{Gamma2}) cannot be neglected.
A more general form of this equation will be discussed a little bit later.

The second of these equations, Eq.\ (\ref{tau_vortex}), 
is the vortex energy-momentum tensor
$\tau^{\mu\nu}_{\rm vortex}$. 
As it should be, it is symmetric and satisfies the condition (\ref{taucond1}).
Moreover, in the non-relativistic limit (when $u^0 \approx 1$ and $u^i\ll 1$)
its time components $\tau^{i0}$ coincide with the energy-density current ${\pmb q}$
(see equation 16.35 in the monograph by Khalatnikov \cite{khalatnikov00}), 
while its spatial components coincide with the non-relativistic vortex stress tensor
(the last term in the right-hand side of Eq.\ \ref{Piik}).
To demonstrate the latter property
it is instructive to rewrite Eq.\ (\ref{tau_vortex})
in terms of the vector $H^{\mu}$. 
One can verify that
\begin{equation}
\tau^{\mu\nu}_{\rm vortex} =
\Gamma \, H^2 g^{\mu\nu} - \Gamma \, H^{\mu}H^{\nu}
+ \Gamma \, H_{\delta} \left( \mathfrak{F}^{\nu\delta}u^{\mu}+\mathfrak{F}^{\mu\delta}u^{\nu}
- H^\delta u^{\mu} u^{\nu}\right),
\label{taumunu} 
\end{equation}
where $H=(H_{\mu}H^{\mu})^{1/2}$ (see also Eq.\ \ref{H2}) and 
$\mathfrak{F}^{\mu\nu} = 1/2 \, \epsilon^{\mu\nu\gamma\delta} F_{\gamma \delta}$ 
is the tensor dual to the vorticity tensor $F^{\mu\nu}$. 
In the non-relativistic limit 
the spatial part of this tensor equals 
$\tau^{ik}_{\rm vortex} \approx \Gamma (H^2 \, \delta^{ik}- H^i H^k)$ [$i,k=1,\,2,\,3$]
and indeed reduces to the non-relativistic expression (see Eq.\ \ref{Piik}),
because in this limit 
${\pmb H} \approx m\, {\rm curl}\, {\pmb V}_{\rm s}=m\, {\pmb \omega}$ 
in the laboratory frame and $\Gamma=\lambda/(mH)=\lambda/(m^2 \omega)$. 

Now, if we 
allow
for the dissipation in the system,
$\tau^{\mu\nu}$ 
will acquire a dissipative correction $\tau^{\mu\nu}_{\rm diss}$, 
so that $\tau^{\mu\nu}=\tau^{\mu\nu}_{\rm vortex}+\tau^{\mu\nu}_{\rm diss}$
and Eq.\ (\ref{entropy2b}) can be rewritten as
\begin{eqnarray}
\partial_{\mu}S^{\mu}&=& \frac{u^{\nu}F_{\mu\nu}}{T} \, \left[Y w^{\mu} 
+ \partial_{\alpha}(\Gamma\, \bot^{\mu \gamma}\bot^{\alpha	 \delta} 
\, F_{\gamma\delta})\right] 
-\partial_{\mu}\left(\frac{u_{\nu}}{T}\right) \, \tau^{\mu\nu}_{\rm diss},
\label{entropy22}
\end{eqnarray}
where $S^{\mu} \equiv Su^{\mu}-u_{\nu}\tau^{\mu\nu}_{\rm diss}/T$ 
is the entropy current density.
Following the consideration of Sec.\ \ref{neglected},
let us introduce the four-vectors 
$f^{\mu}\equiv u^{\nu}F_{\mu\nu}/(\mu n)$ [cf. Eq.\ \ref{f01}]
and $W^{\mu}$,%
%
\footnote{It is interesting to note that 
the ``current density'' $\widetilde{j}^\mu$ defined as 
$\widetilde{j}^\mu \equiv  n u^{\mu}+n W^{\mu}=j^{\mu}
+\partial_{\alpha}(\Gamma\, \bot^{\mu \gamma}\bot^{\alpha	 \delta} 
\, F_{\gamma\delta})$
is conserved, $\partial_{\mu} \widetilde{j}^\mu=0$, 
because $\partial_{\mu}\partial_{\alpha}
(\Gamma\, \bot^{\mu \gamma}\bot^{\alpha	 \delta} \, F_{\gamma\delta})=0$
due to antisymmetry of $F_{\gamma \delta}$.
}
%
\begin{equation}
W^{\mu} \equiv \frac{1}{n}\, \left[Y w^{\mu} 
+ \partial_{\alpha}(\Gamma\, \bot^{\mu \gamma}\bot^{\alpha	 \delta} 
\, F_{\gamma\delta})\right], 
\label{Wmunew}
\end{equation}
and assume that, 
in the comoving frame [$u^{\mu}=(1,\,0,\, 0,\, 0)$], 
the vector $f^{\mu}$ depends only on ${\pmb W}$ and $F_{\mu\nu}$
(see the corresponding discussion after Eq.\ \ref{W1} 
in Sec.\ \ref{neglected}).
Then, positive definiteness 
of the right-hand side of Eq.\ (\ref{entropy22}) means 
independent satisfaction of two conditions,
\begin{eqnarray}
W^{\mu} f_{\mu} &\geq& 0, \quad {\rm and}
\label{posit1}\\
-\frac{1}{T} \,\, \partial_{\mu}u_{\nu} \,\, \tau^{\mu\nu}_{\rm diss}
+\frac{1}{T^2} \,\, \partial_{\mu}T \,\, u_{\nu} \,\, \tau^{\mu\nu}_{\rm diss} &\geq& 0.
\label{posit2}
\end{eqnarray}
The first condition allows us to determine $f^{\mu}$, 
which has the same form as in Eq.\ (\ref{f2}), 
but with $W^{\mu}$ given by Eq.\ (\ref{Wmunew}).%
%
\footnote{Clark's analysis \cite{clark63} of the non-relativistic HVBK-hydrodynamics 
shows that, generally, there can be six independent kinetic coefficients 
instead of three coefficients $\alpha$, $\beta$, and $\gamma$, introduced in Ref.\ \cite{clark63}.
The same consideration applies also to our expression for $f^{\mu}$, 
which is not the most general one (but equivalent to that 
of Ref.\ \cite{khalatnikov00}, see Appendix~\ref{nonrel_limit}).}
%
With this new $f^{\mu}$, the superfluid equation
acquires the same form (\ref{sflrot1}) as in the previous section
[note also that Remark 2 of Sec.\ \ref{neglected} remains 
fully applicable as well].
The second condition allows us to specify the dissipative correction 
$\tau^{\mu\nu}_{\rm diss}$.
This correction can be found in the same way as it was done in Ref.\ \cite{gusakov07};
it includes standard viscous and thermal conduction terms, 
and is presented 
(together with the viscous correction $\varkappa_{\rm diss}$) 
in Appendix \ref{full_hydro}, 
where a complete set of relativistic HVBK-equations is given.

The hydrodynamic equations obtained here fully describe
dynamics of superfluid liquid in the system with vortices
and are equivalent, in the non-relativistic limit,
to the ordinary HVBK-hydrodynamics (see Appendix \ref{nonrel_limit}).

\vspace{0.2 cm}
\noindent
%
{\bf Remark 1.}
There is another, less general, 
way of deriving the tensor $\tau^{\mu\nu}_{\rm vortex}$
by direct averaging
of the ``microscopic'' tensor $T^{\mu\nu}$ (see Eq.\ \ref{Tmunu3}) 
over a volume containing large amount of vortices.
It can be shown that the results of both approaches coincide
(see, in particular, Appendix \ref{micro}, where the
spatial part of the tensor $\tau^{\mu\nu}_{\rm vortex}$ is obtained in this way).

\vspace{0.2 cm}
\noindent
%
{\bf Remark 2.}
Zero-temperature limit of the hydrodynamics described above
can be obtained if we put $T=0$, $S=0$, and $Y=n/\mu$ 
(the latter condition is the relativistic analogue
of the condition $\rho_{\rm s}=\rho$ valid at $T=0$).
Since there are no thermal excitations at $T=0$ 
(except in the vortex cores), 
we also need to specify what we mean by
``the normal-liquid velocity'' $u^{\mu}$, 
which does not have a direct physical meaning in this limit.
In the non-relativistic theory 
the correct superfluid equation valid at $T=0$
will be obtained if we put 
${\pmb V}_{\rm n}={\pmb V}_{\rm s}+(1/\rho) \, {\rm curl \, \lambda {\pmb {\rm e}}}$
(see Appendix \ref{HVBK}, where the same notations are used).
This velocity coincides with the vortex velocity ${\pmb V}_{\rm L}$ (see Eq.\ \ref{Vl}).
The relativistic generalization of this expression 
can be written as
\begin{equation}
u^{\mu}=\frac{m}{\mu} \, V_{({\rm s})}^\mu
+\frac{1}{n} \,\bot^{\mu}_{\nu} \,\, \partial_{\alpha} \left( \Gamma\, \bot^{\nu \gamma}\bot^{\alpha	 \delta} \, F_{\gamma\delta}\right),
\label{relatgen}
\end{equation}
which should be considered as an {\it implicit}
%
\footnote{It is implicit because the right-hand side 
of Eq.\ (\ref{relatgen}) also depends on $u^{\mu}$.
}
%
definition of $u^{\mu}$.
It satisfies the three conditions:

($i$) First, it is easily checked that with this definition 
$u^{\mu}$ is correctly normalized, 
$u_{\mu} u^{\mu}=-1$.

($ii$) Second, one can demonstrate that, with the definition (\ref{relatgen})
one has  
$u^{\nu} W^{\mu} F_{\mu\nu}=0$ 
(see Eq.\ \ref{Wmunew} where $W^{\mu}$ is defined),
i.e., the system entropy remains constant
(see Eq.\ \ref{entropy2b} with $\tau^{\mu\nu}=\tau^{\mu\nu}_{\rm vortex}$).

($iii$) Finally, one can verify that the right-hand side of the superfluid equation
(\ref{sflrot1}) vanishes in view of the expression (\ref{relatgen}), 
which implies
\begin{equation}
u^{\nu}F_{\mu\nu}=0.
\label{sflX}
\end{equation}
One sees (see Remark 3 in Sec.\ \ref{neglected}) 
that, as in the non-relativistic case, 
the vortex velocity coincides with $u^{\mu}$ at $T=0$.
Formula (\ref{sflX}) is the new superfluid equation valid at $T=0$;
$u^{\mu}$ in this equation 
can (in principle) be found by solving equation (\ref{relatgen})
and 
should be considered as a function of $V_{({\rm s})}^{\mu}$. 

\vspace{0.2 cm}
\noindent
%
{\bf Remark 3.}
It can be shown, that the energy-momentum conservation, 
$\partial_{\mu} T^{\mu\nu}=0$, 
which is a superfluous equation in the system with the only one independent velocity 
field $V_{({\rm s})}^{\mu}$, 
is automatically satisfied provided that (\ref{sflX}) holds true.
The resulting system of zero-temperature relativistic HVBK-equations
is thus self-consistent.

\vspace{0.2 cm}
\noindent
%
{\bf Remark 4.}
It would be interesting to compare the zero-temperature version of the relativistic HVBK-hydrodynamics
discussed here
with the results available in the literature.
However, as it is argued in Appendix \ref{incons}, 
we have 
strong concerns 
about self-consistency/validity 
of the existing formulations \cite{lk82, cl95}
of such hydrodynamics.
Thus, no such comparison will be made in the present paper.

\section{Conclusions}
\label{concl}

We have generalized the non-relativistic 
Hall-Vinen-Bekarevich-Khalatnikov (HVBK) hydrodynamics \cite{hv56,bk61}
to the relativistic case. 
The corresponding equations are summarized in Appendix \ref{full_hydro}.
The main difference of the proposed hydrodynamics from the formulations of Refs.\
\cite{lk82, kl82, cl95}
is that it accounts for the presence
of thermal excitations (i.e., is valid at $T\neq0$)
and allows for the interaction between 
the normal and superfluid liquid components (mutual friction).

As a by-product of our work we demonstrate that the previous zero-temperature
formulations of the relativistic vortex hydrodynamics \cite{lk82,cl95}
are internally inconsistent (see Appendix \ref{incons}) and should be modified.

The most natural application of the relativistic HVBK-hydrodynamics
formulated here is to
neutron stars, which are relativistic objects
whose cores are composed of various baryon species (neutrons, protons, etc.) 
that can be in superfluid/superconducting state.
However, to directly apply this hydrodynamics to neutron stars
one should first generalize it to the case of superfluid mixtures
as well as to allow for the possible presence of the magnetic field and 
the related topological defects -- Abrikosov vortices.
These issues were successively addressed, in the non-relativistic framework,
in Refs.\ \cite{ml91, mendell91a, mendell91b, gas11}.
The relativistic generalization of the corresponding equations to the superfluid/superconducting
mixtures without an external magnetic field is straightforward;
the formulation of the full system of magneto-hydrodynamic equations
is more complicated. 
We continue to work in this direction
and hope to present the first results soon.

\begin{acknowledgments}
	
	I am very grateful 
	to Elena Kantor 
	and Vasiliy Dommes 
	for careful reading of an earlier draft 
	of this manuscript and many valuable comments and suggestions.
	Some of the preliminary results of Sec.\ \ref{neglected} 
	were presented by E.M. Kantor at the conference 
	``Electromagnetic Radiation from Pulsars and Magnetars''
	(Zielona Gora, Poland, 2012; see Ref.\ \cite{kg12}).
	This study was supported by the Russian Science Foundation 
	(grant number 14-12-00316).

\end{acknowledgments}

\appendix

\section{HVBK-hydrodynamics}
\label{HVBK}

We present here the main equations of 
dissipative Hall-Vinen-Bekarevich-Khalatnikov hydrodynamics.  
We refer to Refs.\ \cite{hv56, hall60, bk61, khalatnikov00, holm01, donnelly05} for more details.
Hydrodynamic equations in the presence of vortices take the form 
($i$, $k=1$, $2$, $3$)
\begin{eqnarray}
\partial_t \rho+{\rm div} \, {\pmb j} &=& 0,
\label{rhoeq}\\
\partial_t  j^i+{\partial_k} \Pi^{ik} &=& 0,
\label{jeq}\\
\partial_t S+{\rm div} \, S \, {\pmb V}_{\rm n} &=& \frac{R}{T},
\label{Seq}\\
\partial_t {\pmb V}_{\rm s} + ({\pmb V}_{\rm s}{\pmb \nabla}){\pmb V}_{\rm s} 
+ {\pmb \nabla} \left( \breve{\mu}-\frac{1}{2}
\left| {\pmb V}_{\rm s}-{\pmb V}_{\rm n} \right|^2 
\right) &=& {\pmb F},
\label{sfleq1}\\
d E_0 = \breve{\mu} \, d\rho + T \, d S + \rho_{\rm s}({\pmb V}_{\rm s}-{\pmb V}_{\rm n}) \, 
d \left( {\pmb V}_{\rm s} - {\pmb V}_{\rm n}\right) &+& dE_{\rm vortex}.
\label{2ndlawApp}
\end{eqnarray}
and consist of, respectively, continuity equation,
momentum conservation, entropy generation equation, superfluid equation, 
and the second law of thermodynamics.
Here $\rho=m n$ is the density;
$m$ is the particle mass;
$\rho_{\rm s}$ is the superfluid density;
${\pmb j}$ is the mass current density; 
$\Pi^{ik}$ is the stress tensor;
${\pmb V}_{\rm n}$ and $V_{\rm s}$ 
are the normal and superfluid velocities, respectively;
$R$ is the dissipative function; and
${\pmb F}$ is a force to be specified below.
Further, $\breve{\mu}$ is the non-relativistic chemical potential;
in the nonrelativistic limit the chemical potential $\mu$, 
introduced in Sec.\ \ref{Sec2}, 
is related to $\breve{\mu}$ by the formula
$\breve{\mu}=(\mu-mc^2)/m$;
$E_0$ is the non-relativistic energy density 
as measured in the inertial frame moving with the velocity ${\pmb V}_{\rm n}$.
Finally, the last term in Eq.\ (\ref{2ndlawApp})
is responsible for the vortex contribution to the energy density
and is approximately given by \cite{khalatnikov00, holm01}
\begin{equation}
dE_{\rm vortex} = \hat{E}_{\rm vortex} \, d N_{\rm vortex},
\label{dEvortex}
\end{equation}
where
\begin{equation}
\hat{E}_{\rm vortex} = \rho_{\rm s} \frac{\varkappa^2}{4 \pi} \, {\rm ln}\frac{b}{a}
\label{Ekin}
\end{equation}
is the vortex (kinetic) energy per unit length
and
\begin{equation}
N_{\rm vortex}=\frac{\omega}{\varkappa}
\label{Nvort}
\end{equation}
is the area density of vortices.
In Eqs.\ (\ref{Ekin}) and (\ref{Nvort})
${\pmb \omega} \equiv {\rm curl \, {\pmb V}_{\rm s}}$;
$\varkappa = 2 \pi \hbar/(s \, m)$ 
($s=1$ for Bose- and $s=2$ for Fermi-superfluids);
$a$ is the radius of a vortex core;
$b = 1/(\pi N_{\rm vortex})^{1/2}=\varkappa^{1/2}/(\pi \omega)^{1/2}$
is the quantity of the order of the intervortex distance.
Taking into account these definitions, 
Eq.\ (\ref{dEvortex}) can be rewritten as
\begin{equation}
dE_{\rm vortex} = \lambda \, d\omega,
\label{dEvortex2}
\end{equation}
where 
\begin{equation}
\lambda \equiv \rho_{\rm s} \frac{\varkappa}{4 \pi} \, {\rm ln}\frac{b}{a} = 
\rho_{\rm s} \frac{\varkappa}{4 \pi} \, {\rm ln}\frac{\varkappa^{1/2}}{a \pi^{1/2} \omega^{1/2}}.
\label{lambda}
\end{equation}

Hydrodynamic Eqs.\ (\ref{rhoeq})--(\ref{2ndlawApp}) should be 
supplemented by the expressions for 
${\pmb j}$, $\Pi^{ik}$, ${\pmb F}$, and $R$.
Ignoring the thermal diffusivity and viscosity terms 
(which have the standard form,
like in the vortex-free case \cite{bk61, khalatnikov00}),
one has
\begin{eqnarray}
{\pmb j} &=& \rho_{\rm s} {\pmb V}_{\rm s}+\rho_{\rm n} {\pmb V}_{\rm n},
\label{j2}\\
\Pi^{ik} &=& P \delta^{ik} 
+\rho_{\rm s} \, V_{\rm s}^i V_{\rm s}^k + \rho_{\rm n} \, V_{\rm n}^i V_{\rm n}^k
+ \left(\lambda \omega \, \delta^{ik}-\lambda \frac{\omega^i \omega^k}{\omega} \right),
\label{Piik}\\
{\pmb F} &=& -{\pmb \omega}\times \left({\pmb V}_{\rm n}-{\pmb V}_{\rm s} \right)
+ \alpha \, {\pmb \omega}\times \left( {\pmb j}-\rho{\pmb V}_{\rm n}+ {\rm curl} \, \lambda {\pmb {\rm e}} \right)
\nonumber\\
&+& \beta \,  {\pmb {\rm e}} \times \left[  {\pmb \omega}\times \left({\pmb j}-\rho{\pmb V}_{\rm n}+ {\rm curl} \, \lambda {\pmb {\rm e}}\right)\right] \,
\nonumber\\
&-& \gamma \, {\pmb {\rm e}} \, \left[ {\pmb \omega} \left( {\pmb j}-\rho{\pmb V}_{\rm n} 
+ {\rm curl} \, \lambda {\pmb {\rm e}}\right) \right],
\label{F}\\
R &=& - \left[{\pmb F}+ {\pmb \omega}\times \left( {\pmb V}_{\rm n}-{\pmb V}_{\rm s} \right)\right] 
\left({\pmb j}-\rho{\pmb V}_{\rm n}+ {\rm curl} \, \lambda {\pmb {\rm e}} \right).
\label{R}
\end{eqnarray}
Here $\rho_{\rm n}=\rho-\rho_{\rm s}$ is the normal density;
${\pmb {\rm e}} \equiv {\pmb \omega}/\omega$ is the unit vector along 
${\pmb \omega}={\rm curl} \, {\pmb V}_{\rm s}$;
$P=-E_0 + \breve{\mu} \rho+TS$ is the pressure;
$\alpha$, $\beta$, and $\gamma$ are kinetic coefficients 
describing interaction of vortices with the normal liquid component
(mutual friction). 
The term in Eq.\ (\ref{F}), 
depending on $\alpha$, is non-dissipative, 
as opposed to the terms proportional to $\beta$ and $\gamma$.
The coefficients $\beta$ and $\gamma$ 
should be positive in order for the dissipative function $R$
to be positive-definite, $\beta>0$ and $\gamma>0$.
HVBK-equations, described above, deserve a few remarks.

\vspace{0.2 cm}
\noindent
%
{\bf Remark 1.}
Following \cite{ab76, holm01, putterman74}, 
the second law of thermodynamics (\ref{2ndlawApp}) 
is written in a reference frame where the normal liquid component 
is at rest, ${\pmb V}_{\rm n}=0$.
This is in contrast with Refs.\ \cite{bk61, khalatnikov00} where it is written 
in a reference frame 
of a superfluid component, ${\pmb V}_{\rm s}=0$ 
(see, e.g., Ref.\ \cite{holm01} for a more detailed discussion).
As a result, definitions of chemical potential and energy density 
are slightly different in Refs.\ \cite{bk61,khalatnikov00}. 
Namely, it can be shown
that their chemical potential $\breve{\mu}_{\rm Kh}$ 
and the energy density $E_{0 \, {\rm Kh}}$
are related to our $\breve{\mu}$ and $E_0$ by the formulas \cite{holm01} 
\begin{eqnarray}
\breve{\mu}_{\rm Kh} &=& \breve{\mu}-\frac{1}{2}({\pmb V}_{\rm s}-{\pmb V}_{\rm n})^2,
\label{mu1}\\
E_{0 \, {\rm Kh}} &=& E_0 + \frac{1}{2}\, \rho \, ({\pmb V}_{\rm s}-{\pmb V}_{\rm n})^2
-\rho_{\rm s}\, ({\pmb V}_{\rm s}-{\pmb V}_{\rm n})^2.
\label{EEE}
\end{eqnarray}
At the same time, it is easy to verify that 
the pressure in both approaches is the same, $P_{\rm Kh}=P$.
Because the superfluid equation (\ref{sfleq1}) depends on $\breve{\mu}$,
it (formally) differs from the corresponding equation of Refs.\ \cite{bk61, khalatnikov00},
which is expressed through $\breve{\mu}_{\rm Kh}$.

\vspace{0.2 cm}
\noindent
%
{\bf Remark 2.}
As follows from Eqs.\ (\ref{Ekin}) and (\ref{lambda}),  
$\hat{E}_{\rm vortex} \to 0$ and $\lambda \to 0$ at $\hbar \to 0$.
In this limit, corresponding  
to a continuously distributed vorticity 
(like in the ordinary nonsuperfluid hydrodynamics),
contribution of vortices to the total energy and momentum of the liquid can be neglected
(but the ``mutual friction'' terms in Eq.\ \ref{sfleq1}, depending on 
$\alpha$, $\beta$, and $\gamma$, will generally survive).
The situation when one can set $\lambda=0$ 
in all equations described above is common; 
the hydrodynamic equations in this limit are often used, e.g., 
in modelling superfluid dynamics of rotating neutron stars \cite{lm00,ly03}.

\vspace{0.2 cm}
\noindent
%
{\bf Remark 3.}
In the absence of a (generally weak) longitudinal force, $\gamma =0$,
superfluid equation (\ref{sfleq1}) can be rewritten in an elegant way \cite{khalatnikov00}.
Taking the curl of this equation, one obtains
\begin{equation}
\partial_t {\pmb \omega}={\rm curl}({\pmb V}_{\rm L}\times{\pmb \omega}),
\label{rotomega}
\end{equation}
where 
\begin{equation}
{\pmb V}_{\rm L}={\pmb V}_{\rm n} - \alpha \, \left( {\pmb j}-\rho{\pmb V}_{\rm n}+ {\rm curl} \, \lambda {\pmb {\rm e}}\right)
- \beta \, {\pmb {\rm e}} \times \left( {\pmb j}-\rho{\pmb V}_{\rm n}+ {\rm curl} \, \lambda {\pmb {\rm e}} \right).
\label{Vl}
\end{equation}
Equation (\ref{rotomega}) describes translation of the vector ${\pmb \omega}$
with the velocity of the vortex lines ${\pmb V}_{\rm L}$.

Two extreme regimes of motion 
of the vortex lines are of interest.
Assume that it is possible to neglect the terms 
depending on $\lambda$ and $\beta$
in Eqs.\ (\ref{rotomega}) and (\ref{Vl}).
Then, in the {\it strong-drag regime} 
${\pmb V}_{\rm L}={\pmb V}_{\rm n}$ and 
vortices are completely entrained 
by the motion of the normal liquid component.
This regime corresponds to $\alpha=0$.
In the {\it weak-drag regime} the situation is opposite.
Interaction with the normal excitations is so weak that 
vortices move with the superfluid component, 
${\pmb V}_{\rm L}={\pmb V}_{\rm s}$.
Equation (\ref{rotomega}) then takes the form of a standard 
vorticity equation of ordinary hydrodynamics,
\begin{equation}
\partial_t {\pmb \omega}={\rm curl}({\pmb V}_{\rm s}\times{\pmb \omega}).
\label{rotomega2}
\end{equation}
As follows from Eq.\ (\ref{Vl}), the weak-drag limit 
is realized if 
$\alpha= -1/\rho_{\rm s}$.

\section{Relativistic HVBK-hydrodynamics: summary of results}
\label{full_hydro}

Here we present the full system of hydrodynamic equations
which reduces to HVBK-hydrodynamics in the non-relativistic limit.
For the reader's convenience, this appendix is made self-contained.

The main ingredients of the relativistic superfluid HVBK-hydrodynamics
are the four-velocity of thermal excitations $u^{\mu}$, 
normalized by the condition $u_{\mu}u^{\mu}=-1$, 
and the four-vector $w^{\mu}$,
which is defined by Eq.\ (\ref{jmuapp}) (see below).
This four-vector is 
orthogonal to $u^{\mu}$,
\begin{equation}
u_{\mu} w^{\mu} =0,
\label{uwapp}
\end{equation}
and is related to the superfluid velocity $V_{({\rm s})}^{\mu}$
by the formula
\begin{equation}
V_{({\rm s})}^{\mu} = \frac{w^{\mu}+(\mu+\varkappa_{\rm diss})u^{\mu}}{m}, 
\label{Vsmuapp}
\end{equation}
where $m$ is the bare particle mass; $\mu$ is the relativistic chemical potential;
and $\varkappa_{\rm diss}$ is the viscous dissipative correction 
to be specified below (see Eq.\ \ref{kappaapp}).
Another important parameter of this hydrodynamics is the vorticity tensor,
\begin{equation}
F_{\mu\nu}=m \left[\partial_{\mu}V_{({\rm s})\, \nu}-\partial_{\nu}V_{({\rm s})\, \mu} \right].
\label{Fmunuapp}
\end{equation}

The relativistic HVBK-hydrodynamics consists of 
the particle and energy-momentum conservations,
\begin{eqnarray}
\partial_{\mu}j^{\mu} &=&0,
\label{jmueqapp}\\
\partial_{\mu}T^{\mu\nu}&=&0,
\label{Tmunueqapp}
\end{eqnarray}
the second law of thermodynamics 
[note that all the thermodynamic quantities
are measured in the comoving frame, where $u^{\mu}=(1,\,0,\,0,\,0)$], 
\begin{equation}
d \varepsilon = \mu \, dn + T \, dS + \frac{Y}{2} \,  d \left( w_{\mu} w^{\mu} \right) 
+\frac{\lambda}{m} \, dH,
\label{2ndlaw2B}
\end{equation}
and the superfluid equation
\begin{equation}
u^\nu F_{\mu\nu}= \mu n \, f_\mu.
\label{sfleqapp}
\end{equation}

In Eqs.\ (\ref{jmueqapp})--(\ref{sfleqapp})
$n$, $T$, and $S$ are the number density, temperature, and entropy density,
respectively;
$Y$ is the relativistic analogue of the superfluid density \cite{ga06,gusakov07};
$\lambda$ has the same meaning as the corresponding quantity 
of the nonrelativistic HVBK-hydrodynamics (see Appendices \ref{HVBK} and \ref{evortex});
and
\begin{equation}
H=\sqrt{\frac{1}{2} \perp^{\mu\eta} \perp^{\nu\sigma} F_{\mu\nu}  F_{\eta\sigma}},
\label{H2app}
\end{equation}
where $\perp^{\mu\nu}=g^{\mu\nu}+u^{\mu}u^{\nu}$. 
Further, $f^{\mu}$ equals
\begin{equation}
f^{\mu} = \alpha\perp^{\mu\nu} F_{\nu\lambda} \, W_{\delta} \perp^{\lambda\delta}
+ \frac{\beta-\gamma}{H}  \perp^{\mu\eta} \perp^{\nu\sigma}  F_{\eta\sigma} F_{\lambda\nu}  
\, W_{\delta} \perp^{\lambda\delta}
+\gamma  H \, \, W_{\delta} \perp^{\mu\delta},
\label{f2app}
\end{equation}
where $\alpha$, $\beta$, and $\gamma$ are the mutual friction parameters
(the same as in the non-relativistic HVBK-hydrodynamics, see Appendix \ref{HVBK});
and 
\begin{equation}
W^\mu \equiv \frac{1}{n} \left[Y w^{\mu} 
+ \partial_{\alpha}(\Gamma\, \bot^{\mu \gamma}\bot^{\alpha\delta} 
\, F_{\gamma\delta})\right]
\label{Wapp}
\end{equation}
with $\Gamma \equiv \lambda/(mH)$.

It remains to specify the particle current density $j^{\mu}$ 
and the energy-momentum tensor $T^{\mu\nu}$ in Eqs.\ (\ref{jmueqapp}) and (\ref{Tmunueqapp}),
\begin{eqnarray}
j^{\mu} &=& n u^\mu + Y w^\mu,
\label{jmuapp}\\
T^{\mu\nu} &=& (P+\varepsilon) u^{\mu} u^\nu + P g^{\mu\nu} + 
Y \left( w^{\mu} w^{\nu} + \mu u^{\mu} w^{\nu} + \mu u^{\nu} w^{\mu} \right)+\tau^{\mu\nu}_{\rm vortex}
+ \tau^{\mu\nu}_{\rm diss}.
\label{Tmunuapp}
\end{eqnarray}
Here $P=-\varepsilon+\mu n + TS$ is the pressure;
and $\tau^{\mu\nu}_{\rm vortex}$ is the vortex contribution to $T^{\mu\nu}$,
\begin{equation}
\tau^{\mu\nu}_{\rm vortex}
=\Gamma\,   \perp_{\delta \alpha} F^{\mu\delta} F^{\nu \alpha}
- \Gamma \, u^{\mu} u^{\nu} u^{\gamma} u_{\beta} \, F^{\alpha \beta} F_{\alpha \gamma}.
\label{tauvortexapp}
\end{equation}
Finally, the dissipative corrections 
$\varkappa_{\rm diss}$ and $\tau^{\mu\nu}_{\rm diss}$ 
in Eqs.\ (\ref{Vsmuapp}) and (\ref{Tmunuapp}) are given by \cite{gusakov07}
\begin{eqnarray}
\varkappa_{\rm diss} &=&  
- \xi_3 \, \partial_{\mu} \left( Y w^{\mu} \right)
- \xi_4 \, \partial_{\mu} u^{\mu},
\label{kappaapp}\\
\tau^{\mu \nu}_{\rm diss} &=&  
- \kappa \, \left( \perp^{\mu \gamma} \, u^{\nu} 
+ \perp^{\nu \gamma} \, u^{\mu} \right) 
\left(  \partial_{\gamma} T + T u^{\delta} \,
\partial_{\delta} u_{\gamma} \right)
\nonumber \\
&-& \eta \, \perp^{\mu \gamma} \, \perp^{\nu \delta} \,\, 
\left( \partial_{\delta} u_{\gamma} + \partial_{\gamma} u_{\delta}  
- \frac{2}{3} \,\, g_{\gamma \delta} \,\, 
\partial_{\varepsilon} u^{\varepsilon}  \right) 
\nonumber \\
&-& \xi_1 \, \perp^{\mu \nu} \, \partial_{\gamma} \left( Y w^{\gamma} \right) 
- \xi_2 \, \perp^{\mu \nu} \, \partial_{\gamma} u^{\gamma}.
\label{tau_munu1app}
\end{eqnarray}
In these equations 
$\kappa$ and $\eta$ are, respectively, the thermal conductivity 
and shear viscosity coefficients; 
$\xi_1$,$\ldots$,$\xi_4$ are the bulk viscosity coefficients ($\xi_1=\xi_4$; 
$\xi_1^2\leq\xi_2\xi_3$; $\kappa$, $\eta$, $\xi_2$, $\xi_3\geq 0$).

\section{Superfluid equation in the non-relativistic limit}
\label{nonrel_limit}

Equations of the relativistic HVBK-hydrodynamics
are summarized in Appendix \ref{full_hydro}. 
Our aim here will be to demonstrate that the ``superfluid'' equation 
(Eq.\ \ref{sfleqapp}) of this hydrodynamics reduces to its non-relativistic 
counterpart (\ref{sfleq1}) in the nonrelativistic limit.
In what follows we use dimensional units.
In these units Eq.\ (\ref{sfleqapp}) becomes
\begin{equation}
u^\nu F_{\mu\nu}= \frac{\mu n}{c^3} \, f_\mu.
\label{sfleqapp2}
\end{equation}

Spatial components 
of Eq.\ (\ref{sfleqapp2}) can be rewritten as 
($i$, $j=1$, $2$, $3$)
\begin{equation}
u^0 F_{i0}+u^{j}F_{ij}=\frac{\mu n}{c^3} \, f_i,
\label{spatialNR}
\end{equation}
or, in view of (\ref{Fmunuapp}),
\begin{equation}
m u^0 \left[\partial_i V_{({\rm s})\, 0}-\partial_0 V_{({\rm s})\, i}\right]  + 
m u^{j} \left[\partial_i V_{({\rm s})\, j}-\partial_j V_{({\rm s})\, i}\right]
=\frac{\mu n}{c^3} \, f_i, \quad {\rm or}
\label{spatialNR2}
\end{equation}
\begin{equation}
\partial_i V_{({\rm s})\, 0} = \partial_0 V_{({\rm s})\, i}
-\frac{u^{j}}{u^0} \left[\partial_i V_{({\rm s})\, j}-\partial_j V_{({\rm s})\, i}\right]
+\frac{\mu n}{m c^3 \, u^0} f_i.
\label{spatialNR3}
\end{equation}
On the other hand, 
it follows from the orthogonality condition (\ref{uwapp}), that
\begin{equation}
u^{\mu} V_{({\rm s})\, \mu}=-\frac{\mu}{m c}  \,\, \Rightarrow \,\,
V_{({\rm s})\, 0}  = -\frac{\mu}{m c \, u^0} - \frac{u^{j}}{u^0} \, V_{({\rm s})\, j}.
\label{orthNR}
\end{equation}
where we made use of the definition (\ref{Vsmuapp}),
which takes the form (in dimensional units 
and neglecting the dissipative correction $\varkappa_{\rm diss}$)
\begin{equation}
V_{({\rm s})}^{\mu} = \frac{w^{\mu}+\mu u^{\mu}}{m c}.
\label{VsmuNR}
\end{equation}

Substituting Eq.\ (\ref{orthNR}) into (\ref{spatialNR3}),
one obtains
\begin{eqnarray}
\partial_0 V_{({\rm s})\, i}
-\frac{u^{j}}{u^0} \left[\partial_i V_{({\rm s})\, j}-\partial_j V_{({\rm s})\, i}\right]
+\frac{\mu n}{m c^3 \, u^0} f_i= -\partial_i \left(\frac{\mu}{m c \, u^0} \right)
- \partial_i \left(\frac{u^{j}}{u^0} \, V_{({\rm s})\, j} \right).
\label{spatialNR4}
\end{eqnarray}
Now, introducing the non-relativistic chemical potential 
$\breve{\mu}=(\mu-mc^2)/m$ (see Appendix \ref{HVBK})
and taking into account that $u^{\mu}$ 
is expressed through the velocity ${\pmb V}_{\rm n}$ 
of the normal component as
\begin{equation}
u^{\mu}=\left(\frac{1}{\sqrt{1-{\pmb V}_{\rm n}^2/c^2}},\, 
\frac{{\pmb V}_{\rm n}}{c\sqrt{1-{\pmb V}_{\rm n}^2/c^2}} \right),
\label{umuNR}
\end{equation}
one arrives at the following equation, 
valid at $|{\pmb V}_{\rm n}|$, $|{\pmb V}_{s}| \ll c$ 
[we recall that ${\pmb V}_{\rm s} \equiv(V_{({\rm s})}^1,\,V_{({\rm s})}^2,\,V_{({\rm s})}^3)$],
\begin{eqnarray}
\partial_t {\pmb V}_{\rm s} + {\rm curl} {\pmb V}_{\rm s} \times {\pmb V}_{\rm n}
+ {\pmb \nabla}\left[ \breve{\mu} 
- \frac{1}{2} {\pmb V}_{\rm n}^2+{\pmb V}_{\rm n}{\pmb V}_{\rm s}\right]
= -\frac{\mu n}{m c^2} \, f^i,
\label{spatialNR5}
\end{eqnarray}
or, taking into account that ${\pmb \nabla} ({\pmb V}_{\rm s}^2)/2=
({\pmb V}_{\rm s}{\pmb \nabla}){\pmb V}_{\rm s} 
- {\rm curl}{\pmb V}_{\rm s}\times {\pmb V}_{\rm s}$
and $\mu \approx m c^2$,
\begin{eqnarray}
\partial_t {\pmb V}_{\rm s} + ({\pmb V}_{\rm s}{\pmb \nabla}){\pmb V}_{\rm s} 
+ {\pmb \nabla}\left[ \breve{\mu} 
- \frac{1}{2} \left| {\pmb V}_{\rm s}-{\pmb V}_{\rm n}\right|^2\right]
=-{\rm curl}{\pmb V}_{\rm s}\times \left({\pmb V}_{\rm n}-{\pmb V}_{\rm s}\right) 
-n \, f^i.
\label{spatialNR6}
\end{eqnarray}
This equation is very similar to Eq.\ (\ref{sfleq1}), 
but to draw a final conclusion we need also to analyze the spatial part $f^i$
of the four-vector $f^{\mu}$ (see Eq.\ \ref{f2app}). 
In the non-relativistic limit it is given by Eq.\ (\ref{fifi}),
\begin{eqnarray}
{\pmb f}=-\alpha\, m \, [{\rm curl} \,  {\pmb V}_{\rm s} \times {\pmb W}]
-\beta\, m\,  {\pmb {\rm e}} \times[{\rm curl} 
\, {\pmb  V}_{\rm s} \times {\pmb W}]+
\gamma\, m \,  {\pmb {\rm e}} ({\pmb W}\,{\rm curl} \, {\pmb V}_{\rm s}),
\label{fifiNR}
\end{eqnarray}
where ${\pmb W}$ is the spatial part of the four-vector $W^{\mu}$ (see Eq.\ \ref{Wapp}),
which is, in the dimensional form, 
\begin{equation}
W^\mu = \frac{1}{n} \left[c \, Y w^{\mu} 
+ \partial_{\alpha}(\Gamma\, \bot^{\mu \gamma}\bot^{\alpha\delta} 
\, F_{\gamma\delta})\right].
\label{WNR}
\end{equation}
The spatial component of the first term here equals, 
in the nonrelativistic limit, 
$c\, Y w^{\mu}= \rho_{\rm s}({\pmb V}_{\rm s}-{\pmb V}_{\rm n})/m$
(see Eq.\ \ref{VsmuNR} and note that 
$Y=\rho_{\rm s}/(m^2 c^2)$ at $c \to \infty$ \cite{ga06,gusakov07}).

The last term in Eq.\ (\ref{WNR}), 
which equals $\partial_\alpha (\Gamma\,O^{\mu\alpha})$, 
can be rewritten as (see Eq.\ \ref{Omunu} for the definition of $O^{\mu\alpha}$ 
and the footnote \ref{footnote8})
\begin{equation}
\partial_\alpha (\Gamma\,O^{\mu\alpha})=\partial_\alpha 
\left( \frac{\Gamma}{2} \, \epsilon^{\delta\eta\mu\alpha} \, u_{\eta} \,
\epsilon_{\delta a b c} \, u^a \, F^{bc}\right).
\label{termNR}
\end{equation}
In the nonrelativistic limit the only terms here that survive
are those with $\eta=0$ and $a=0$. 
Because both $\partial_{\alpha} u_0$ 
and $\partial_{\alpha} u^0$ are of the order of $1/c^2$ 
(see Eq.\ \ref{11}), 
one may treat $u_{\eta}$ and $u^a$
in Eq.\ (\ref{termNR}) as constants 
($u^0\approx1$ and $u_0\approx-1$).
In this way one finds ($i=1$, $2$, $3$), 
\begin{equation}
\partial_\alpha (\Gamma\,O^{i\alpha}) = \epsilon^{i\alpha\delta} \, 
\partial_{\alpha}\left( \Gamma \, \frac{1}{2} \, \epsilon^{\delta b c} F^{bc} \right)
= m \, {\rm curl}(\Gamma\, {\rm curl {\pmb V}_{\rm s}})
= \frac{{\rm curl}(\lambda\,{\pmb {\rm e}})}{m},
\label{termNR2}
\end{equation}
where we employed Eq.\ (\ref{Gamma2}), 
and used the fact that $H=m \, |{\rm curl} {\pmb V}_{\rm s}|$ (see Eq.\ \ref{Hmu}).
Returning then to the vector ${\pmb W}$, one can write
\begin{equation}
{\pmb W} = \frac{1}{mn} \left[\rho_{\rm s}({\pmb V}_{\rm s}-{\pmb V}_{\rm n}) 
+{\rm curl}(\lambda\,{\pmb {\rm e}}) \right].
\label{vectorWNR}
\end{equation}
Substituting now Eqs.\ (\ref{fifiNR}) and (\ref{vectorWNR})
into Eq.\ (\ref{spatialNR6}), 
one verifies that it coincides 
with the superfluid equation (\ref{sfleq1})
of nonrelativistic HVBK-hydrodynamics.

\section{Energy of a relativistic vortex 
and the expression for $d \varepsilon_{\rm vortex}$}
\label{evortex}

Let us consider a homogeneous system without vortices.
We assume that all the thermodynamic parameters, as well as the velocities of normal
$u^{\mu}$ and superfluid $V_{({\rm s})0}^{\mu}=\partial^{\mu}\phi_0/m$ 
components are constants in time and space.
All the quantities related to this (unperturbed) 
system will be denoted by the subscript ``0''.

In what follows we shall work in the coordinate frame in which $u^{\mu}=(1, 0, 0, 0)$ 
[hereafter, the normal-liquid coordinate frame].
In that frame the quantity $w^{\mu}_{(0)}=\partial^{\mu}\phi_0-\mu_0 u^{\mu}$ 
can be represented as
$w^{\mu}_{(0)}=(0,\, \partial^i \phi_0)$ 
on account of Eq.\ (\ref{wu}) 
[$i=1$, $2$, $3$ is the spatial index]. 
Correspondingly, the energy density $T^{00}_{(0)}$ 
is given simply by 
$\varepsilon_0=\varepsilon_0(S_0, \, n_0,\, w_{\mu \, (0)}w^{\mu}_{(0)})$ 
(see Eq.\ \ref{Tmunu3}), 
which is generally a function 
of the entropy density $S_0$, the number density $n_0$, 
and the scalar $w_{\mu \, (0)}w^{\mu}_{(0)}$.

Now let us {\it adiabatically} perturb the system by creating a straight vortex,
assuming that the total number of particles remains unchanged.
Denoting the 
correction due to the vortex as $\phi_{\rm V}$,
one finds for the perturbed system (in the normal-liquid frame): 
$w^{\mu}=(0,\,  \partial^{i} \phi_0 + \partial^{i} \phi_{\rm V})$ 
and $T^{00}=\varepsilon(n,\, S,\, w_{\mu} w^{\mu})$.
The vortex energy can be defined as the difference between 
the energies of the perturbed and unperturbed systems,
\begin{equation}
E_{\rm V}=\int dV \, (T^{00}-T^{00}_{(0)})=
\int d V \, [\varepsilon(S,\, n,\, w_{\mu}w^{\mu})
-\varepsilon_{\rm 0}(S_{\rm 0},\, n_{\rm 0},\, w_{\mu \,(0)}w^{\mu}_{(0)})],
\label{Ev}
\end{equation}
where the integration is performed over the system volume $V$.
As it will be clear from the subsequent consideration, the main contribution 
to $E_{\rm V}$ comes from the region far from the vortex, 
where $S({\pmb r})$, $n({\pmb r})$, and $w_{\mu}({\pmb r})w^{\mu}({\pmb r})$
only weakly deviate from, respectively, 
$S_0$, $n_0$, and $w_{\mu \, (0)}w^{\mu}_{(0)}$.
Consequently, one can expand the function under the integral in Eq.\ (\ref{Ev}) 
and present $E_{\rm V}$ as
\begin{eqnarray}
E_{\rm V}
&\approx& \int dV \frac{\partial \varepsilon}{\partial S} \, [S({\pmb r})-S_0]+
\int dV \frac{\partial \varepsilon}{\partial n} \, [n({\pmb r})-n_0]+
\int dV \frac{\partial \varepsilon}{\partial(w_{\mu}w^{\mu})} \, [w^{\mu}w_{\mu}-w^{\mu}_{(0)}w_{\mu \, (0)}]
\nonumber\\
&\approx& T_0 \int dV \, [S({\pmb r})-S_0]+
\mu_0 \int dV \, [n({\pmb r})-n_0]+
\frac{Y_0}{2} \int dV  \, [w^{\mu}w_{\mu}-w^{\mu}_{(0)}w_{\mu \, (0)}],
\label{Ev2}
\end{eqnarray}
where in the second equality use has been made 
of the second law of thermodynamics (\ref{2ndlawXXX}).
Since the total entropy and particle number in the perturbed and unperturbed systems 
are the same by construction%
%
\footnote{
This is not strictly true because formula (\ref{Ev2}) does not include 
integration over the volume in the immediate vicinity of the vortex core,
where the hydrodynamic approach is not applicable. 
However, the entropy
and the number of particles contained in that volume 
are small (proportional to the radius $a$ of the vortex core squared),
hence their contribution to the total entropy 
and particle number can be neglected.
},
%
the first two integrals vanish, so that
\begin{equation}
E_{\rm V} \approx
\frac{Y_0}{2} \int dV  [w^{\mu}w_{\mu}-w^{\mu}_{(0)}w_{\mu \, (0)}]=
\frac{Y_0}{2} \int dV \, [\partial^{i} \phi_{\rm V} \, \partial_{i} \phi_{\rm V}
+2 \, \partial^{i}\phi_{\rm V} \, \partial_{i}\phi_{\rm 0}].
\label{Ev3}
\end{equation}
Because $\partial_i \phi_0$ is constant
and $\partial^{i}\phi_{\rm V}$ is symmetric (see Eq.\ \ref{phiV} below),
the contribution into the integral from the second term in Eq.\ (\ref{Ev3})
vanishes
and we finally arrive at the following formula for $E_{\rm V}$,
\begin{equation}
E_{\rm V} \approx
\frac{Y_0}{2} \int dV  \, \partial^{i} \phi_{\rm V} \, \partial_{i} \phi_{\rm V}.
\label{Ev4}
\end{equation}
In the non-relativistic limit (\ref{Ev4}) reduces to the standard
expression for the vortex energy,
\begin{equation}
E_{\rm V}=\frac{\rho_{\rm s0}}{2}\int dV \, {\pmb V}_{\rm s V}^2,
\label{Evnonrel}
\end{equation}
if we note that the superfluid velocity induced by the vortex, ${\pmb V}_{\rm s  V}$, is 
related to the scalar $\phi_{\rm V}$ by the condition ${\pmb V}_{\rm s  V}={\pmb \nabla}\phi_{\rm V}/m$
and that in the non-relativistic limit the superfluid density $\rho_{\rm s 0}=m^2 Y_0$. 

To take an integral in Eq.\ (\ref{Ev4})
one needs to specify $\partial^i \phi_{\rm V}$.
If the straight vortex is at rest in the normal-liquid frame and $\partial^i \phi_0=0$
then, as follows from the symmetry arguments 
(see Eq.\ \ref{int22}),  
it induces the velocity field given simply by%
%
\footnote{To obtain Eq.\ (\ref{phiV})
we choose a circle in 3D centered at the vortex line
as the integration contour in Eq.~(\ref{int22}).
\label{15d}}
%
\begin{equation}
\partial^i \phi_{\rm V}=\frac{{\rm {\pmb e}}_{\varphi}}{s r}, 
\label{phiV}
\end{equation}
where ${\rm {\pmb e}}_{\varphi}$ is the unit vector in the azimuthal direction 
($\varphi$ is the polar angle) 
and $s=1$ or 2 is the quantity defined after Eq.\ (\ref{int22}).
In reality, however, we deal with a non-stationary problem:
the vortex can move with some velocity in the normal-liquid frame
and $\partial^i \phi_0$ does not necessary vanish.
In the non-relativistic theory it is argued (e.g., Ref.\ \cite{khalatnikov00}) 
that Eq.\ (\ref{phiV}) remains a good approximation 
for $\partial^i \phi_{\rm V}$ even in this case.
The latter result can be extended to the fully relativistic case
if we assume that the background superfluid velocity $\partial^i \phi_{0}$
(and hence the vortex velocity) are much smaller than the speed of light $c$ 
in {\it the normal-liquid frame}. 
In an arbitrary frame this requirement means that the difference 
between the spatial components of normal and background superfluid velocities 
should be much smaller than the 
speed of light $c$.
This condition is not very restrictive and, for example, 
is satisfied in the superfluid matter of neutron stars, 
where superfluidity is destroyed long before
the velocity difference becomes comparable to $c$ \cite{gk13}.

Substituting (\ref{phiV}) into (\ref{Ev4}) and performing an integration, one arrives
at the following expression for the vortex energy per unit length
(we suppress the subscript ``0'' from here on), 
\begin{equation}
\hat{E}_{\rm V}=\frac{\pi Y}{s^2} \,\, {\rm ln} \frac{b}{a},
\label{Ev5}
\end{equation}
where $a$ is the radius of the vortex core and 
$b$ is an ``external'' radius of the order 
of the inter-vortex spacing (as in the non-relativistic theory).
The radius $b$ is related to the number of vortices $N_{\rm V}$ 
per unit area by the standard formula (cf.\ Ref.\ \cite{khalatnikov00}),
\begin{equation}
\pi b^2 = \frac{1}{N_{\rm V}}.
\label{bb}
\end{equation}
On the other hand, 
as follows from Eq.\ (\ref{int22}) and the Stokes' theorem 
(see also Eq.\ \ref{int33}), 
$N_{\rm V}$ is related to the smooth-averaged vorticity 
(defined in the normal-liquid frame) 
by the expression 
\begin{equation}
N_{\rm V}= \frac{s \, m \, |\varepsilon^{ijk} \partial_j V_{({\rm s}) \, k}|}{2\pi}.
\label{bbb}
\end{equation}
To obtain this formula an integration is performed over the surface whose boundary 
is the contour specified in the footnote \ref{15d}.

Using Eqs.\ (\ref{bb}) and (\ref{bbb}),
one obtains the following expression 
for the vortex energy density $\varepsilon_{\rm vortex}$,
\begin{equation}
\varepsilon_{\rm vortex}=\frac{\hat{E}_{\rm V}}{\pi b^2}=
 \frac{m Y}{2s} \,{\rm ln} \left(\frac{b}{a}\right) \,|\varepsilon^{ijk} \partial_j V_{({\rm s}) \, k}|
 \equiv \lambda \, |\varepsilon^{ijk} \partial_j V_{({\rm s}) \, k}|,
\label{evv}
\end{equation}
where we introduced the parameter $\lambda \equiv [m Y/(2s)] \, {\rm ln}(b/a)$,
which only weakly (logarithmically) depends on $b$ 
(and, as a consequence, 
weakly depends on $|\varepsilon^{ijk} \partial_j V_{({\rm s}) \, k}|$).
Note that $\varepsilon_{\rm vortex}$ is the quantity which is determined, 
by definition, in the normal-liquid frame.
It is thus a Lorentz-invariant quantity
and it is useful to rewrite it in an explicitly Lorentz-invariant form.
One can do this
with the help of the four-vector $H^{\mu}$ (see Eq.\ \ref{Hmu}),
\begin{equation}
\varepsilon_{\rm vortex}= \frac{\lambda}{m} \sqrt{H^{\mu}H_{\mu}}.
\label{evv2}
\end{equation}
Consequently, the differential of this energy density 
due to a variation of $H^{\mu}$ is given by
\begin{equation}
d\varepsilon_{\rm vortex}=
\frac{\lambda}{2mH} \, d(H_{\mu}H^{\mu}),
\label{evv3}
\end{equation}
where $H=\sqrt{H^{\mu}H_{\mu}}$ and we neglected, 
as in the non-relativistic theory, 
the dependence of $\lambda$ on $H^{\mu}$.
This formula coincides with the expression (\ref{dEvort}) 
for $d \varepsilon_{\rm vortex}$
used in the text.

\vspace{0.2 cm}
\noindent
%
{\bf Remark 1.}
Presence of vortices not only adds an additional term
(\ref{evv3}) to the second law of thermodynamics (\ref{2ndlaw2})
but also renormalizes the particle chemical potential $\mu$,
which is now approximately given by (see Eq.\ \ref{evv2})
\begin{equation}
\mu = \mu_{\rm old}+\frac{\partial \varepsilon_{\rm vortex}}{\partial n}
= \mu_{\rm old}+\frac{H}{m}\, \frac{\partial \lambda}{\partial n},
\label{munow}
\end{equation}
where $\mu_{\rm old}$ is the chemical potential in the absence of vortices.

\section{Spatial part of the tensor $\tau^{\mu\nu}_{\rm vortex}$ from the
microscopic averaging procedure}
\label{micro}

Here we briefly demonstrate how to obtain 
the spatial part of the tensor $\tau^{\mu\nu}_{\rm vortex}$
from the ``microscopic'' tensor $T^{\mu\nu}$ (see Eq.\ \ref{Tmunu3}). 
In the comoving frame 
[i.e., in the frame in which $u^{\mu}=(1,0,0,0)$]
the spatial components of the tensor $T^{\mu\nu}$
equal
\begin{equation}
T^{ij}=P\,  g^{ij}+Y \, w^{i} w^{j}
=P \, g^{ij} + Y\, \partial^{i} \phi \, \partial^{j} \phi.
\label{PijE}
\end{equation}
In the system with vortices $\phi=\phi_0+\phi_{\rm V}$ 
(the notations are the same as in Appendix \ref{evortex}).
Assume that we have a bunch of vortices with locally constant density,
which are directed along the axis~$z$.
Let us introduce a notion of the ``Wigner-Seitz cell''
-- a cylinder of radius $b$ surrounding each vortex line.
We then average the tensor $T^{ij}$ out over one such Wigner-Seitz cell.
Since 
we neglect interaction between vortices, 
the
neighbouring vortices 
``do not interfere'' when averaging Eq.\ (\ref{PijE}). 
The result can be written as
\begin{equation}
\langle T^{ij}\rangle = \frac{1}{\pi b^2 \Delta z} \, \int dV \, T^{ij}
= \langle P \rangle g^{ij} 
+ \langle Y \rangle \, \partial^{i} \phi_0 \, \partial^{j} \phi_0  
+ \langle Y  \, \partial^{i} \phi_{\rm V} \, \partial^{j} \phi_{\rm V} \rangle,
\label{PijE2}
\end{equation}
where angle brackets mean averaging over the Wigner-Seitz cell;
$\Delta z \sim b$ is a height of cylinder 
(the actual value of $\Delta z$ is not important);
and $dV$ is the volume element.
Note that
the main contribution to $\langle Y \rangle$
comes from the 
region 
far 
from the vortex core, 
where $Y$ can be considered as constant
(the dependence of $Y$ on $\partial^{i}\phi_{\rm V}$ is weak).
This means that
$\langle Y \rangle \approx Y_0$, where $Y_0$ 
is the value of $Y$ at a distance $\sim b$ from the vortex 
(or, equivalently, the value of $Y$ in the system without vortices; see Appendix \ref{evortex}).
Similarly, one can also replace $Y$ with $Y_0$ when taking other averages.

The cross-terms $\langle Y \, \partial^i \phi_{0} \, \partial^j \phi_{\rm V} \rangle$
and $\langle Y \, \partial^j \phi_{0} \, \partial^i \phi_{\rm V} \rangle$
in Eq.\ (\ref{PijE2})
vanish 
on account of the
symmetry of the problem. 
Clearly, the only ``interesting'' (non-standard) 
contribution to $\langle T^{\mu\nu} \rangle$
comes from the last term in Eq.\ (\ref{PijE2}),  
which can be identified as the vortex tensor, i.e.,
$\tau^{ij}_{\rm vortex}
=\langle Y  \, \partial^{i} \phi_{\rm V} \, \partial^{j} \phi_{\rm V} \rangle$.
To find this tensor, let us write $\partial^i \phi_{\rm V}$
in Cartesian coordinates ($x$, $y$, $z$) using Eq.\ (\ref{phiV}),
\begin{eqnarray}
\partial^{x} \phi_{\rm V} &=& -\frac{{\rm sin}\varphi}{sr},
\label{VxVyE1}\\
\partial^{y} \phi_{\rm V} &=& \frac{{\rm cos}\varphi}{sr},
\label{VxVyE22}\\
\partial^{z} \phi_{\rm V} &=& 0,
\label{VxVyE}
\end{eqnarray}
where the
axes $x$ and $y$ are located in the plane perpendicular to the axis $z$ 
(and all the three axes cross at the vortex line).
Using Eqs.\ (\ref{VxVyE1})--(\ref{VxVyE})
it is easily verified that the only non-zero components of the vortex tensor 
are $\tau^{xx}_{\rm vortex}$ and $\tau^{yy}_{\rm vortex}$;
they are given by
\begin{eqnarray}
\tau^{xx}_{\rm vortex} &=& \tau^{yy}_{\rm vortex}
=\langle Y  \, \partial^{x} \phi_{\rm V} \, \partial^{x} \phi_{\rm V} \rangle
=\frac{1}{\pi b^2 \Delta z} \, \int dz\,  r dr \, d\varphi \,Y \, 
 \frac{{\rm sin}^2\varphi}{s^2 r^2}
 \approx \frac{Y_0}{s^2 b^2} \, {\rm ln} \frac{b}{a}
\nonumber\\ 
 &=&\lambda \, |\epsilon^{ijk} \partial_j V_{({\rm s}) \, k}|,
\label{VxVyE2}
\end{eqnarray}
where $\lambda$ is defined by the same formula as in Appendix \ref{evortex}
and we used Eqs.\ (\ref{bb}) and (\ref{bbb}) to obtain the last equality.
Making 3D rotation, $\tau^{ij}_{\rm vortex}$
can generally be presented as 
(${\pmb H}=m \, {\rm curl} \,{\pmb V}_{\rm s}$)
\begin{equation}
\tau^{ij}_{\rm vortex}=\frac{\lambda}{m} H - \frac{\lambda}{m} \frac{H^i H^j}{H}.
\label{TijvortexE2}
\end{equation}
This tensor exactly coincides with the spatial part of the tensor $\tau^{\mu\nu}_{\rm vortex}$,
written in the comoving frame and
presented in Sec.\ \ref{included} (see Eq.\ \ref{taumunu} there).
Interestingly, the same tensor $\tau^{ij}_{\rm vortex}$ can be determined
from the purely thermodynamic arguments following the method of Ref.\ \cite{ep77}.

\section{Inconsistency of the zero-temperature vortex hydrodynamics of Refs.~\cite{lk82,cl95}}
\label{incons}

Here we shall demonstrate 
that the vortex hydrodynamics of Ref.\ \cite{lk82}
is internally inconsistent and hence the energy-momentum tensor $T^{\mu\nu}$
of that hydrodynamics should be modified.
Since Ref.\ \cite{cl95} obtained the same expression%
%
\footnote{See Appendix B of that reference.}
%
for $T^{\mu\nu}$
(although 
some of its other equations are different 
due to some unexplained reason),
it suffers from the same inconsistency problem.
Thus,
we shall not discuss Ref.\ \cite{cl95} in what follows.
The notations used in this section differ 
from those adopted in other parts of the paper
and coincide with the notations of Ref.\ \cite{lk82}.

Let us consider the formula (77) of Ref.\ \cite{lk82}.
It gives the energy-momentum tensor $T^{\mu\nu}$
for the superfluid liquid with the distributed vorticity 
at $T=0$.
This tensor can be written as
\begin{equation}
T^{\mu}_{\nu}=\frac{c^2}{\mu_0}\frac{\partial \Phi}{\partial \mu_0} \, v^{\mu} v_{\nu}
+\frac{\omega^{\mu} \omega_{\nu}}{\omega} \, \frac{\partial \Phi}{\partial \omega}
- (\Phi-\omega \frac{\partial \Phi}{\partial \omega}) \, \delta^{\mu}_{\nu},
\label{Tmunu2x}
\end{equation}
where $\mu_0$ is the invariant chemical potential;
$\Phi$ is the invariant pressure;
$v^{\mu}$ is the superfluid velocity normalized by the condition 
$v_{\mu} v^{\mu}=\mu_0^2/c^2$ (see equation 54 of Ref.\ \cite{lk82});
and $\delta^{\mu}_{\nu}$ is the Kronecker symbol.
Finally, $\omega=\sqrt{-\omega_{\mu}\omega^{\mu}}$, 
where the four-vector $\omega^{\mu}$ 
is the generalization of ${\rm curl \, {\pmb V}}_{\rm s}$
to the relativistic case; it is given by the formula (74) of Ref.\ \cite{lk82}.
Below we assume that the metric is flat 
and equals $g_{\mu\nu}=(c^2,-1,-1,-1)$ 
(see a formula after equation 27 in Ref.\ \cite{lk82}).

Consider a tensor $T^{\mu\nu}$ in the coordinate frame in which
$v^{\mu}=(\mu_0/c^2,0,0,0)$.
In this frame the time component of 
the four-vector $\omega^0$ vanishes, 
$\omega^0=0$ (see equation 74 of Ref.\ \cite{lk82}),
hence the energy density $\varepsilon$, 
given by the component $T^0_0$ of the tensor $T^{\mu}_{\nu}$, 
equals
\begin{equation}
\varepsilon=T^{0}_{0}=\frac{c^2}{\mu_0} \, \frac{\partial \Phi}{\partial \mu_0} \, \frac{\mu_0^2}{c^2}
- (\Phi-\omega \frac{\partial \Phi}{\partial \omega})
= - \Phi + \mu_0 \, \frac{\partial \Phi}{\partial \mu_0} +\omega \, \frac{\partial \Phi}{\partial \omega}.
\label{epsilonX}
\end{equation}
This expression can be rewritten if one introduces the mass density
$\rho=m_0 n$, where $m_0$ is the mass of a free particle and 
$n$ is the number density.
As follows from the formula (76) of Ref.\ \cite{lk82} 
for $j^{\mu}$ ($j^{\mu}$ is the density of the mass 4-flux), 
in the chosen coordinate frame 
$\rho=j^0=\partial \Phi/\partial \mu_0$,
thus equation (\ref{epsilonX})
can be represented as
\begin{equation}
\varepsilon= - \Phi + \mu_0 \rho + \omega \, \frac{\partial \Phi}{\partial \omega}.
\label{epsilon2}
\end{equation}
This formula seems to be incorrect 
(contradicts other equations of Ref.\ \cite{lk82}).
The simplest way to demonstrate 
this
is to look at the non-relativistic limit 
of the vortex hydrodynamics of Ref.\ \cite{lk82}.
In this limit Eq.\ (\ref{epsilon2}) should reduce to the 
corresponding non-relativistic expression for the energy density
if we set to zero the superfluid velocity ${\pmb V}_{\rm s}$ in 
the latter expression 
(in other words, 
we consider a point in space in which ${\pmb V}_{\rm s}=0$
at some particular moment of time).

The non-relativistic expression for the energy density
was obtained in Ref.\ \cite{bk61} and is presented in the monograph
by Khalatnikov \cite{khalatnikov00} on p.~101, 
\begin{equation}
E_0=-P+TS+\breve{\mu}_{\rm Kh} \, \rho + ({\pmb V}_{\rm n}-{\pmb V}_{\rm s}){\pmb j}_0,
\label{E}
\end{equation}
where $E_0$ is the nonrelativistic energy density measured in the frame in which ${\pmb V}_{\rm s}=0$
(not to be confused with $E_0$ from Appendix \ref{HVBK}!); 
$P$ is the pressure defined as in Ref.\ \cite{khalatnikov00}; 
$\breve{\mu}_{\rm Kh}$ is the non-relativistic chemical potential, 
which equals $\breve{\mu}_{\rm Kh}=\partial E_0/\partial \rho$ 
(in the monograph by Khalatnikov \cite{khalatnikov00} 
this potential is denoted by $\mu$); 
$j_0=\rho_{\rm n}({\pmb V}_{\rm n}-{\pmb V}_{\rm s})$; 
$\rho_{\rm n}$ is the normal density.
In our case we have $T=0$, thus ${\pmb j}_0=0$
and (\ref{E}) can be rewritten as
\begin{equation}
E_0=-P+\breve{\mu}_{\rm Kh}\, \rho.
\label{E2}
\end{equation}
The non-relativistic energy $E_0$ and the pressure $P$
are related to their relativistic counterparts
by the formulas
\begin{eqnarray}
\varepsilon &=& E_0+\rho c^2,
\nonumber\\
\Phi &=& P,
\label{pe}
\end{eqnarray}
where the last equality is needed to reproduce 
the correct non-relativistic 
stress tensor $\Pi^{ik}$ from Eq.\ (\ref{Tmunu2x}).
Moreover, $\mu_0$ can be presented as 
$\mu_0= c^2 + \delta \mu_0$, where $\delta \mu_0$ is a small correction.
As a result, one obtains that the formula (\ref{epsilon2})
transforms to the form
\begin{equation}
E_0=-P + \delta \mu_0 \rho + \omega \frac{\partial \Phi}{\partial \omega}.
\label{E3}
\end{equation}
Comparing (\ref{E2}) and (\ref{E3}) one sees that
\begin{equation}
\delta \mu_0 = \frac{1}{\rho} \, \left( \breve{\mu}_{\rm Kh} \, \rho 
- \omega \, \frac{\partial \Phi}{\partial \omega}\right),
\label{conc}
\end{equation}
i.e., $\delta \mu_0 \neq \breve{\mu}_{\rm Kh}$.
In other words, the ``invariant'' chemical potential $\mu_0$ 
is not simply given by the partial derivative $\partial \varepsilon/\partial \rho$,
where $\varepsilon$ is the energy density measured in the frame in which 
$v^{\mu}=(\mu_0/c^2,0,0,0)$.
This is a strange result 
(in which frame is then $\mu_0$ specified
as a derivative of the energy density with respect to the mass density?)
that contradicts, in particular, the non-relativistic superfluid equation
(80) presented in Ref.\ \cite{lk82}.
In order to make the equation (80) of Ref.\ \cite{lk82}
compatible 
with the corresponding equation of nonrelativistic HVBK-hydrodynamics
(written for a point in space in which ${\pmb V}_{\rm s}=0$
at some particular moment of time; see the equation 16.40 
of the monograph \cite{khalatnikov00} by Khalatnikov with 
$\rho_s=\rho$ and $\beta'=\beta=\gamma=0$),
it is necessary to have 
$\nabla v_0=\nabla \breve{\mu}_{\rm Kh}$, 
i.e.,
$\delta \mu_0 = \breve{\mu}_{\rm Kh}$ (since in the chosen reference frame 
$v_0=\mu_0=c^2+\delta\mu_0$),
in contradiction with (\ref{conc}).

We come to conclusion that the vortex hydrodynamics of Ref.\ \cite{lk82} 
(and hence Ref.\ \cite{cl95})
is internally inconsistent: equation (\ref{epsilon2}) is not correct 
(the last term in its right-hand side is superfluous),
which 
means that the
energy-momentum tensor (\ref{Tmunu2x}) 
should be modified.


\end{document}